%% file: ANA-HION-2018-28-PAPER.tex
\documentclass[PAPER, cernpreprint, atlasdraft=false, texlive=2023, USenglish, texmf, orcidlogo]{atlasdoc}
 
\usepackage{placeins}
\usepackage{amsmath}
\usepackage{atlaspackage}
\usepackage{atlasbiblatex}
 
\usepackage{atlasphysics}

\graphicspath{{logos/}{figures/}}
 
\usepackage{ANA-HION-2018-28-PAPER-defs}

\AtlasTitle{Measurements of the suppression and correlations of dijets in Xe+Xe collisions at $\sqrt{s_{NN}}$ = 5.44 TeV}

\AtlasAbstract{
Measurements of the suppression and correlations of dijets is performed using 3~$\mu$b$^{-1}$ of Xe+Xe data at $\sqrt{s_{\mathrm{NN}}} = 5.44$~TeV collected with the ATLAS detector at the LHC\@. Dijets with jets reconstructed using the $R=0.4$ anti-$k_t$ algorithm are measured differentially in jet $p_{\text{T}}$ over the range of 32~GeV to 398~GeV and the centrality of the collisions. Significant dijet momentum imbalance is found in the most central Xe+Xe collisions, which decreases in more peripheral collisions. Results from the measurement of per-pair normalized and absolutely normalized dijet $p_{\text{T}}$ balance are compared with previous Pb+Pb measurements at $\sqrt{s_{\mathrm{NN}}} =5.02$~TeV. The differences between the dijet suppression in Xe+Xe and Pb+Pb are further quantified by the ratio of pair nuclear-modification factors. The results are found to be consistent with those measured in Pb+Pb data when compared in classes of the same event activity and when taking into account the difference between the center-of-mass energies of the initial parton scattering process in Xe+Xe and Pb+Pb collisions. These results should provide input for a better understanding of the role of energy density, system size, path length, and fluctuations in the parton energy loss.
 
}

\AtlasRefCode{HION-2018-28}
 
\PreprintIdNumber{CERN-EP-2023-001}

\AtlasJournal{PRC}
\AtlasJournalRef{\PRC 108 (2023) 024906}
\AtlasDOI{10.1103/PhysRevC.108.024906}

\hypersetup{pdftitle={ATLAS document},pdfauthor={The ATLAS Collaboration}}
 
\begin{document}
\maketitle
\tableofcontents
 
\renewcommand{\xj}{x_{\mathrm{J}}}
 
\section{Introduction}
\label{sec:intro}
A major focus of relativistic heavy-ion physics is to study the quark-gluon plasma (QGP), a hot and dense medium composed of deconfined quarks and gluons. During the initial stages of heavy-ion collisions, hard-scattering interactions between constituents of incoming nuclei may occur. In leading-order calculations in perturbative quantum chromodynamics, two high transverse momenta (\pt) partons (quarks and/or gluons) are produced in these interactions. These partons then fragment and hadronize to form two jets that are oriented back-to-back in azimuth. When traversing the QGP, these jets suffer radiative and collisional energy loss leading to a phenomenon known as jet quenching~\cite{Qin:2015srf,Blaizot:2015lma}. Jet quenching has been observed and quantified in many measurements at RHIC and the LHC (for a recent review see Ref.~\cite{Cunqueiro:2021wls}), but the theoretical understanding of partonic interactions and properties of the QGP is still incomplete.
 
A basic observable quantifying the impact of jet quenching on inclusive jets is the jet nuclear-modification factor,
\RAA~\cite{HION-2017-10,CMS-HIN-13-005,ALICE:2019qyj}. A factor of two suppression of inclusive jet production in central \pbpb\ collisions compared to the
production in \pp\ collisions is seen in measurements of \RAA.
While measurements of inclusive jet suppression contain contributions from jets traversing different
path lengths in the QGP and suffering differently from jet quenching fluctuations, the measurements of dijets provide additional new information.
The increase in the number of strongly asymmetric dijets may be
directly connected with the different path lengths of the two jets in
the medium \cite{Qin:2010mn}.
The compound effect of fluctuations in the vacuum-like fragmentation
pattern and medium related fluctuations in the energy loss may also significantly
contribute to the measured asymmetry \cite{Milhano:2015mng}.

Dijet suppression in \pbpb\ collisions was measured
in terms of the momentum balance~\cite{HION-2010-02,CMS-HIN-10-004,CMS-HIN-11-013,HION-2012-11,HION-2019-02} and the pair nuclear-modification
factor~\cite{HION-2019-02}, which quantified the differences between the suppression of the leading (in transverse momentum) jet and the subleading jet (opposite in
azimuth).

So far, only inclusive charged-hadron suppression was measured in \xexe collisions~\cite{ATLAS:2022kqu,CMS-HIN-18-004,ALICE:2018hza}, while the parton energy-loss measurements involving jets and dijets at the LHC have only been performed in \pbpb collisions. The 2017 LHC \xexe\ run provides a possibility to study jet quenching in collisions of nuclei lighter than Pb, which is attractive for several reasons.
First, the underlying event (UE) in the most central collisions, where the collision geometry is the most symmetric, is smaller in \xexe collisions than in \pbpb collisions. Secondly, the decrease in the number of nucleons or the nuclear radius between Pb and Xe nuclei may affect the amount of jet quenching through a reduction in both the overall energy density and the path lengths traversed by the hard-scattered partons in the medium. Consequently, measurements of dijets in different collision systems should further constrain the impact of the path length, energy density, and fluctuations on the jet quenching. Studying the parton energy loss in a collision system that is smaller than the \pbpb system may also help to predict the energy loss for oxygen--oxygen collisions, which are intended to be performed during LHC Run~3~\cite{Citron:2018lsq}. This paper reports the first measurement of jet suppression for Xe+Xe collisions.
 
The analysis presented here follows closely the techniques that were used to measure dijets in Pb+Pb collisions~\cite{HION-2019-02}. The highest $\pt$ jet in the event, the leading jet, and the second highest $\pt$ jet, the subleading jet are studied. The leading jet transverse momentum, $\ptlead$ is required to be greater than 100~\GeV\ and the subleading jet transverse momentum, $\ptsub$, greater than 32~\GeV.
Both the jets are required to be in the rapidity\footnote{ATLAS uses a right-handed coordinate system with its origin at the nominal interaction point (IP) in the center of the detector and the $z$-axis along the beam pipe. The $x$-axis points from the IP to the center of the LHC ring, and the $y$-axis points upward. Cylindrical coordinates $(r,\phi)$ are used in the transverse plane, $\phi$ being the azimuthal angle around the $z$-axis. The pseudorapidity is defined in terms of the polar angle $\theta$ as $\eta=-\ln\tan(\theta/2)$ and the rapidity is defined in terms of the energy $E$ and $z$-component of the momentum, $p_z$, as $y=1/2 \ln((E+p_z)/(E-p_z))$. Transverse momentum and transverse energy are defined as $\pt = p\sin(\theta)$ and $E_\text{T}=E\sin(\theta)$, respectively. The angular distance between two objects with relative differences $\Delta\eta$ in pseudorapidity and $\Delta\phi$ in azimuth is given by $\Delta R = \sqrt{(\Delta\eta)^2 + (\Delta\phi)^2}$.} region  $|y| < 2.1$.
The subleading jet is required to be on the opposite side in azimuth from the leading jet, which is defined by the condition $\Delta\phi = |\phi_1 - \phi_2| >7\pi/8$. The momentum balance between the leading and subleading jet is quantified by the ratio
$$
\xj = \frac{\ptsub}{\ptlead}.
$$
The distribution of $\xj$ can be normalized by the number of dijets ($N_\mathrm{pair}$),
\begin{equation}
\label{eq:dxja}
\frac{1}{N_\mathrm{pair}} \frac{\dd N_\mathrm{pair}}{\dd\xj},
\end{equation}
which is called \textit{per-pair normalized} \asymmetry distribution. Alternatively, the distribution can be normalized by the effective heavy-ion luminosity,
\begin{equation}
\label{eq:dxjb}
\frac{1}{N_{\mathrm{evt}}} \frac{1}{\TAAavr} \frac{\dd N_\mathrm{pair}}{\dd\xj},
\end{equation}
which is called the \textit{absolutely normalized} \asymmetry distribution, where $\TAAavr$ is the average nuclear-thickness function and $N_\mathrm{evt}$ is the number of minimum bias events in a given centrality interval. The absolutely normalized \asymmetry distribution allows the differences between the yields of dijets with a given dijet \asymmetry in different collision centrality intervals to be directly quantified.
Both the per-pair normalized and absolutely normalized dijet \asymmetry distributions were previously measured in \pbpb collisions~\cite{HION-2019-02}.
The dijet yields in \xexe collisions are also extracted to calculate the pair nuclear-modification factors for leading and subleading jets. Measurements made in \xexe collisions are compared with those obtained for \pbpb collisions.

\section{ATLAS detector and trigger}
 
The ATLAS detector~\cite{PERF-2007-01}  at the LHC covers nearly the full solid angle around the nominal interaction point.
It contains an inner tracking detector (ID) surrounded by a thin superconducting solenoid, electromagnetic and hadronic calorimeters, a zero-degree calorimeter, and a muon spectrometer that incorporates three large
superconducting toroidal magnets. The inner-detector system is immersed in a 2~T axial magnetic
field and provides charged-particle tracking in the range of $|\eta|~<~2.5$ with 2$\pi$ coverage in azimuth. The ATLAS calorimeter system covers the pseudorapidity range of $|\eta|~<~4.9$. In the region of $|\eta|~<~3.2$, the electromagnetic calorimetry is provided by both barrel and endcap high-granularity lead/liquid-argon (LAr) calorimeters, with an additional thin LAr presampler covering $|\eta|~<~1.8$, to enable corrections for energy lost in material upstream of the calorimeters. Hadronic calorimetry is provided
by a steel/scintillating-tile calorimeter that is segmented into three barrel structures with $|\eta|~<~1.7$, and two copper/LAr hadronic endcap calorimeters. To complete the solid angle coverage, forward ($3.2~<~|\eta|~<~4.9$)
copper/LAr and tungsten/LAr calorimeter modules (FCal) are used, optimized for electromagnetic and hadronic measurements, respectively.
 
Data are recorded with a multi-stage trigger system~\cite{TRIG-2016-01}. Events are selected using  hardware-based first-level triggers (L1) implemented in custom-electronics, and then processed by a high-level trigger (HLT) to further reduce the accepted event rate and provide additional purity.
 
An extensive software suite~\cite{ATL-SOFT-PUB-2021-001} is used in data simulation, in the reconstruction and analysis of real and simulated data, in detector operations, and in the trigger and data acquisition systems of the experiment.
 
\section{Data selection and Monte Carlo simulation samples}
The analysis uses data from $^{129}$\ce{Xe}+$^{129}$\ce{Xe} collisions at $\sqrtsnn = 5.44$~\TeV\ collected in 2017 at the LHC with a total integrated luminosity of 3~$\mu$b$^{-1}$. Events were recorded using a combination of two triggers designed to select minimum-bias (MB) collisions, which allows measured jets to be reconstructed with full efficiency.
These triggers require the total transverse energy deposited in the calorimeters at L1 to be more than 4~\GeV\ or if the total transverse energy at L1 is less than 4~\GeV\ then the presence of at least one track reconstructed in the ID is required.

In addition to the trigger selections, events are required to have a reconstructed primary vertex and satisfy criteria that ensure stable detector conditions. A few recorded events ($\approx 0.1\%$) consistent with two Xe+Xe interactions in the same bunch crossing (pileup) are removed based on the tight correlation between
the sum of the total transverse energy in the forward calorimeter ($\sumETFCal$) and the number of reconstructed tracks matched to the primary vertex.
 
The level of overall event activity or ``centrality'', which is indicative of the degree of overlap between the two colliding nuclei, is characterized using
$\sumETFCal$ measured at the electromagnetic scale~\cite{PERF-2011-03}. The Glauber model~\cite{Loizides:2017ack} is used to obtain a correspondence between the $\sumETFCal$ distribution and the fraction of the total inelastic \xexe cross-section, allowing the setting of the centrality percentiles~\cite{HION-2011-01, HION-2011-05}. A Glauber Model analysis was also applied to relate quantiles of the $\sumETFCal$ distribution to geometric properties of the collision such as $\TAAavr$. Centrality intervals in \xexe collisions used in the analysis along with $\TAAavr$ are summarized in Table~\ref{tab:FCalEtXeXe}.
The comparison of \xexe and \pbpb results is performed in the same centrality intervals, covering the centrality range 0--80\% as listed in Table~\ref{tab:FCalEtXeXe}. In addition, the results for both systems are also compared by selecting events with similar activity, quantified by $\sumETFCal$. For this comparison, events in Xe+Xe collisions are selected in intervals of $\sumETFCal$ matching the $\sumETFCal$ intervals in \pbpb collisions that correspond to the \pbpb centrality intervals of 10--20\%, 20--40\% and 40--60\% used in Ref.~\cite{HION-2019-02}. The corresponding \xexe centrality intervals are summarized in Table~\ref{tab:FCalEtXeXeCompare}.
 
\begin{table*}
\caption{The centrality intervals in \xexe collisions and their corresponding $\TAAavr$ values with their respective absolute uncertainties.}
\label{tab:FCalEtXeXe}
\centering
\begin{tabular}{r@{}lr@{}l}
\hline
\multicolumn{2}{c}{Centrality} & \multicolumn{2}{c}{\TAAXeXe\  [mb$^{-1}$]} \\
\hline
0&--10\%  &  $12.38$&$\pm0.08$ \\
10&--20\% &  $7.53$&$\pm0.09$  \\
20&--40\% &  $3.52$&$\pm0.09$  \\
40&--80\% &  $0.630$&$\pm0.036$  \\ \hline
\end{tabular}
\vspace{2mm}
\end{table*}
 
\begin{table*}
\caption{The centrality intervals in \xexe and \pbpb collisions for matching $\sumETFCal$ intervals and respective \TAAavr\ values for \xexe collisions.}
\label{tab:FCalEtXeXeCompare}
\centering
\begin{tabular}{r@{}lr@{}lr@{}lr@{}l}
\hline
\multicolumn{2}{c}{\xexe Cent.} & \multicolumn{2}{c}{\pbpb Cent.} & \multicolumn{2}{c}{\TAAXeXe\ [mb$^{-1}$]} & \multicolumn{2}{c}{$\sumETFCal$ [\TeV]}\\
\hline
0&--7.7\% & 10&--20\% &  $13.05$&$\pm0.08$ & 2.06&--3.00  \\
7.7&--29.9\% & 20&--40\% & $6.45$&$\pm0.09$ & 0.89&--2.06  \\
29.9&--53.2\% & 40&--60\% & $1.81$&$\pm0.07$ & 0.30&--0.89  \\ \hline
\end{tabular}
\vspace{2mm}
\end{table*}
 
Monte Carlo (MC) simulations are used to understand the performance of the ATLAS detector in high occupancy \xexe data samples and to correct the data for detector effects. A sample of $9\times10^6$ \pp\ jet events was generated using \PYTHIA8~\cite{Sjostrand:2014zea} at $\sqrt{s} =5.44$~\TeV\ with the A14 set of tuned parameters~\cite{ATL-PHYS-PUB-2014-021} and the NNPDF23LO parton distribution functions~\cite{Ball:2012cx}. To correctly describe the UE of \xexe collisions, these generated MC events were overlaid onto events from a dedicated sample of minimum-bias \xexe data. The detector response was simulated~\cite{ATL-SOFT-PUB-2021-001} using \textsc{Geant4}~\cite{Agostinelli:2002hh,SOFT-2010-01}.
A \pp\ MC sample of $2.4 \times 10^6$ jet events with the same settings but at $\sqrt{s} =5.02$~\TeV\ was also generated to determine the correction for the difference between the center-of-mass energies in \xexe and \pbpb collisions. In addition, samples of jet events using \HERWIGpp~\cite{Bahr:2008pv} with the UEEE5 tune~\cite{Gieseke:2012ft} and the CTEQ6L1 parton distribution functions~\cite{Pumplin:2002vw} were generated to assess systematic uncertainties.

\section{Jet reconstruction}
 
The jet reconstruction procedure follows that used by ATLAS for previous jet measurements in \pbpb collisions described in Ref.~\cite{HION-2020-09}, including the UE subtraction procedure. Jets are reconstructed using the \antikt algorithm~\cite{Cacciari:2008gp} with radius parameter $R=0.4$ implemented in the \verb=FastJet= software package \cite{Fastjet}. Jets are formed by clustering calorimetric towers of angular size $\Delta\eta \times \Delta\phi = 0.1 \times \pi/32$. The energy in the tower is obtained by summing the energies deposited in calorimeter cells at the electromagnetic energy scale within the tower boundaries. An $\eta$- and $\phi$-dependent UE subtraction is performed for each calorimeter tower within the jet using an iterative procedure, where the background due to the UE is modulated
to account for the effects of hydrodynamic flow~\cite{HION-2016-06}.
Then, jet $\eta$ and $\pt$ dependent correction factors derived from simulations are applied to the measured jet energy to correct for the calorimeter energy response~\cite{PERF-2012-01}. This calibration is followed by a cross-calibration that relates the jet energy scale of jets reconstructed by the procedure outlined above to the jet energy scale in 13~\TeV \pp\ collisions~\cite{ATLAS-CONF-2015-016}.
An additional correction based on in situ studies of jets recoiling against photons, $Z$ bosons, and jets in other regions of the calorimeters is applied~\cite{HION-2018-06}.

Jets are defined at the generator level in the MC sample before detector simulation by applying the \antikt algorithm with $R=0.4$ to stable particles with a proper lifetime greater than 30~ps, but excluding muons and neutrinos, which do not leave significant energy deposits in the calorimeter. After the detector simulation, the generator-level jets are matched to the nearest reconstructed jet within $\Delta R=0.4$.
 
\begin{figure*}
\centering
\includegraphics[width=0.49\textwidth]{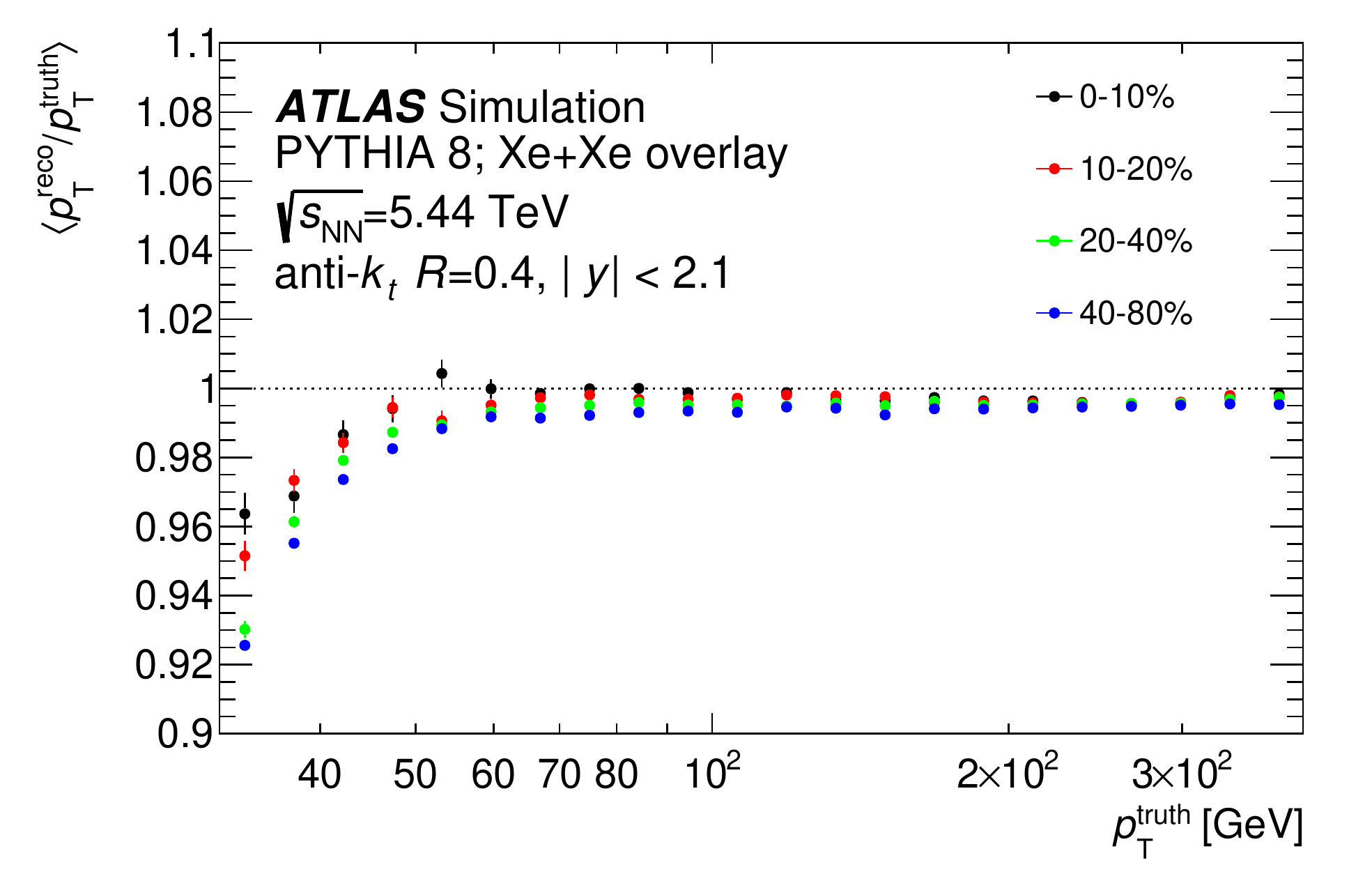}
\includegraphics[width=0.49\textwidth]{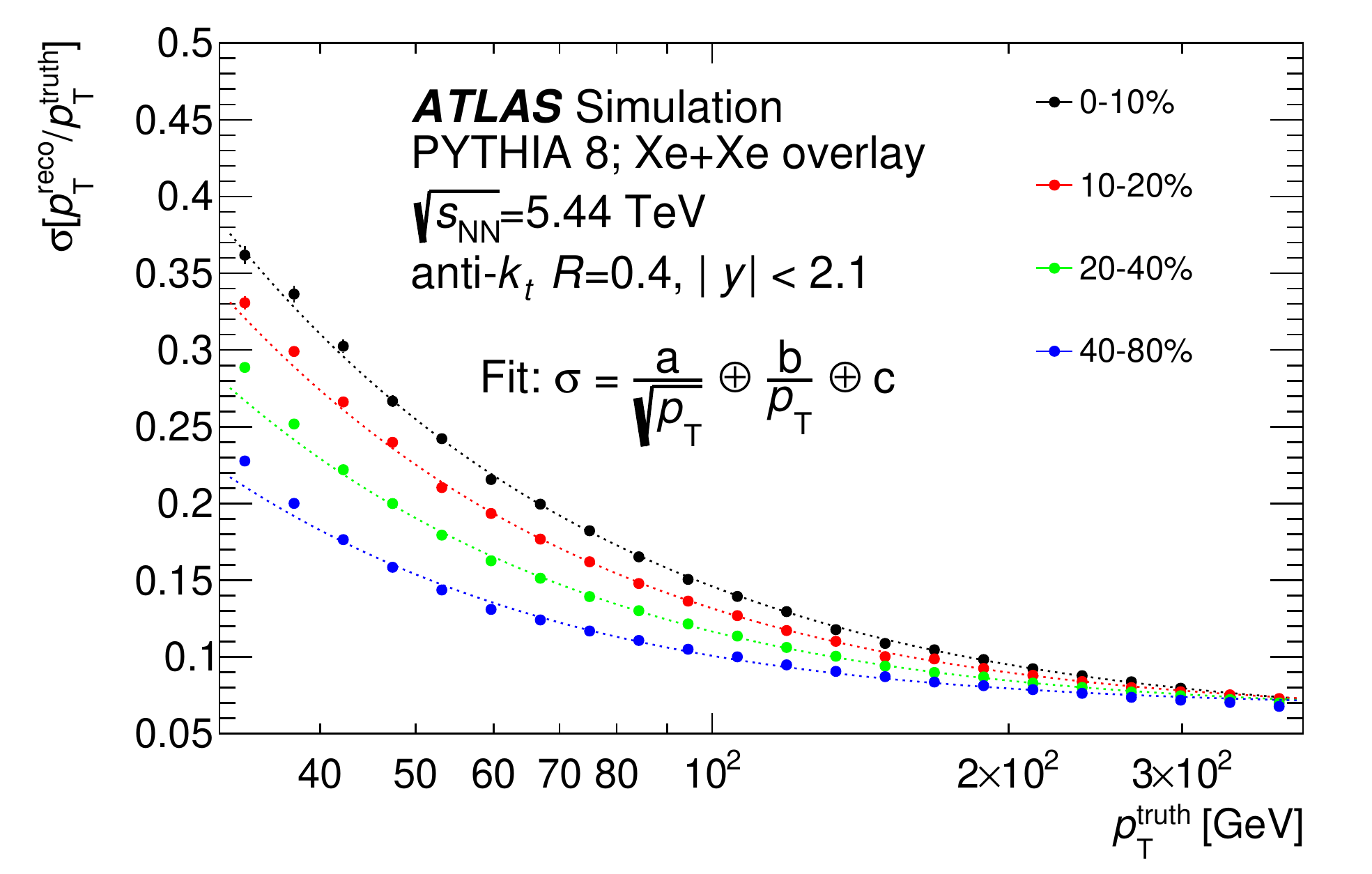}
\caption{
The performance of (left) the JES and (right) the JER for jets with $|y| < 2.1$ evaluated as a function of $\pt^{\text{truth}}$ in different centrality bins. The right plot includes JER fits shown with dashed curves (for details, see the text).
Simulated hard scatter events were overlaid onto events from a dedicated sample of minimum-bias \xexe data.
}
\label{fig:jesjer}
\end{figure*}

The performance of the jet reconstruction is shown in Figure~\ref{fig:jesjer} in terms of the jet energy scale (JES) and jet energy resolution (JER), which correspond to the mean and width of the jet response $(\pt^{\text{reco}} / \pt^{\text{truth}})$, where $\pt^{\text{reco}}$ and $\pt^{\text{truth}}$ are the reconstructed and generator level jet transverse momenta, respectively. The maximum departure of the JES from unity in the inclusive jet sample is 8\% in peripheral collisions (40–80\% centrality) for $\pt^\text{truth}<50$~\GeV. For $\pt^\text{truth}>50$~\GeV\ the JES is consistent with unity within 1\%. A similar performance of the JES is seen in \pbpb collisions~\cite{HION-2019-02}. Contributions to the JER can be factorized into three terms,
$\frac{a}{\sqrt{\pt}} \oplus \frac{b}{\pt} \oplus c$, where the constants $a$, $b$, and $c$ quantify the magnitude of stochastic term, noise term, and constant term, respectively.
The stochastic and constant terms are related to the calorimeter response to the showering process while the noise term constitutes the centrality-dependent part of JER that is driven by the \pt-independent UE fluctuations~\cite{HION-2011-02}.
The JER is largest in 0--10\% central collisions where $b = 11.3 \pm 0.1$~\GeV, which is about 3~\GeV\ smaller than the value of $b$ in 0--10\% central \pbpb collisions but similar to the value of $b$ in 10--20\% central \pbpb collisions~\cite{HION-2017-10}. Constants $a$ and $c$ are found to be consistent between \xexe and \pbpb collisions.
The impact of the small departure of the JES from unity on measured quantities and the impact of the JER is corrected for by the unfolding procedure.

\section{Data analysis}
\label{sec:analysis}
 
The analysis is performed using the same methods as described in Ref.~\cite{HION-2019-02}. Jet pairs are formed from the two highest-$\pt$ jets in the event.
Jet pairs not fulfilling the selection criteria for leading and subleading jet defined in Section~\ref{sec:intro} are not used.
 
The distributions of dijet \asymmetry and the pair nuclear-modification factor are not calculated directly. They are obtained in the final step of the analysis from two-dimensional $\ptonepttwo$ distributions.
The binning of $\ptonepttwo$ distributions follows a logarithmic distribution with 32 intervals between $\pt = 10$~\GeV\ and $\pt = 398$~\GeV.
This binning allows the same intervals to be selected as those used in the previous dijet and inclusive jet measurements in \pbpb collisions~\cite{HION-2019-02,HION-2017-10}.
Distributions are measured in the \pt range of $32-398$~\GeV. Bins with $\pt < 32$~\GeV\ are only used in MC simulation as underflow bins in the unfolding procedure. The maximum value of 398~\GeV\ is determined by the absence of jets with $\pt > 398$~\GeV\ in the data. In total, 14~325 dijets were analyzed in the full centrality range $0-80$\%.
 
The $\ptonepttwo$ distributions are corrected for the background from jet pairs not originating in the same hard process and from spurious jets from fluctuations of the UE\@. This combinatoric background is estimated from the data using the yield of dijets with angular separation $1.0 < \Delta\phi < 1.4$ from the leading jet.
This angular requirement minimizes the contribution from real dijets, which have a maximum at $\Delta\phi = \pi$, and contributions from split jets in the vicinity of $\Delta\phi \approx 0.4$. The background jet yield is then subtracted from the yield of dijets with  $\Delta\phi > 7\pi/8$.
This correction is largest in the most central collisions and at low \pt. In the 0--10\% centrality region, for $\ptlead > 100$~\GeV\ and $32 <\ptsub <50$~\GeV, it subtracts up to 15\% of the dijets while for all other \pt and centrality bins the correction subtracts less than 4\% of the dijets.
The presence of background jets may create a situation where such a  jet has higher \pt\ than the real sub-leading jet. This background jet would then be falsely identified as the sub-leading jet and this would cause oversubtraction leading to inefficiency. The background-subtracted $\ptonepttwo$ distributions are corrected for this effect. The efficiency correction is estimated from the per-event rate of inclusive jets in data using the same procedure as described in Ref.~\cite{HION-2012-11}. After the background of size $B$ is subtracted from the raw yields $N^\mathrm{raw}$, the efficiency $\epsilon$ is applied using the formula: $N^\mathrm{corr} = (N^\mathrm{raw}-B)/\epsilon$, where $N^\mathrm{corr}$ is the final yield that goes into the unfolding. The efficiency correction is largest in the most central collisions at the lowest $\ptsub$ values, where it reaches $3\%$. For subleading jets with $\ptsub>50$~\GeV\ the efficiency correction is smaller than $1\%$.
 
After the combinatoric background-subtraction and efficiency correction, the $\ptonepttwo$ distributions are unfolded for the detector response using the Bayesian unfolding implemented in the \verb=RooUnfold= package~\cite{Adye:2011gm,DAgostini:1994fjx}. The four-dimensional response matrices are filled symmetrically in reconstruction- and generator-level $\ptonepttwo$ to include the possibility that the leading and subleading jets are swapped due to resolution effects. Generator-level jets entering the response matrix satisfy the same $y$ and $\Delta\phi$ conditions, but extend the minimum $\pt$ to lower values ($\ptlead > 20$~\GeV\ and $\ptsub > 10$~\GeV) to account for the migration of jets in and out of the kinematic fiducial region. The unfolding procedure also corrects the jet reconstruction inefficiency when the reconstructed jet pair is lost due to resolution effects but the corresponding generator-level jet pair exists. The response matrices are reweighted in generator-level, \ptonepttwo by smooth ratios of the \ptonepttwo\ distributions in data to those in the reconstructed MC sample such that the \ptonepttwo\ distributions in the response matrices better represent those in the data.
 
The number of iterations in the unfolding is chosen to be three for all the centrality intervals, which optimizes the balance between the statistical uncertainty and systematic bias introduced by the shapes of the distributions used to construct the response matrix. The statistical uncertainty is estimated by performing 100 unfoldings where each bin in input data and each bin in the response matrix are varied separately according to their corresponding statistical uncertainties. The standard deviation of these is evaluated in each bin and used as the estimate of the statistical uncertainty. The two values, one from the statistical uncertainty in the data and one from the statistical uncertainty in the response matrix are summed in quadrature to obtain the total statistical uncertainty in the unfolded distributions.
 
The unfolded $\ptonepttwo$ distributions are used to calculate the resulting one-dimensional
dijet \asymmetry distributions defined in Equations~(\ref{eq:dxja}) and (\ref{eq:dxjb}). The unfolded $\ptonepttwo$ is mapped to the region $\ptsub \leq \ptlead$ and diagonally sliced to project the $\xj$ distribution as described in Ref.~\cite{HION-2019-02}.
The $\xj$ bin boundaries are defined by the previously discussed logarithmic binning in transverse momentum.
The $\xj$ distributions measured in \xexe collisions can be directly compared with distributions measured in \pbpb collisions. This comparison may be affected by the difference between the cross-sections, due to the different center-of-mass energies of the initial hard process scattering, in \xexe and \pbpb collisions. To quantify this difference, the factor \cxj\ is calculated as a ratio of \PYTHIA[8] $\xj$ distributions in 5.44 \TeV\ \pp collisions to the same quantity in simulated 5.02 \TeV\ \pp collisions,
\begin{equation}
\label{eq:cxj}
\cxj = \frac{
1/N  ~ \dd N^{\mathrm{pair}}_{\mathrm{\PYTHIA8}}(pp,5.44~\mathrm{\TeV})/ \dd \xj
}{
1/N  ~ \dd N^{\mathrm{pair}}_{\mathrm{\PYTHIA8}}(pp,5.02~\mathrm{\TeV})/ \dd \xj
}.
\end{equation}
The normalization $N$ is $N_\mathrm{pair}$ and $N_\mathrm{evt} \TAAavr$ for per-pair normalized and absolutely normalized distributions, respectively. The \cxj\ factor can be used to scale the $\xj$ distributions measured in \pbpb data. For the absolutely normalized $\xj$ distributions, the magnitude of \cxj\ ranges from 1.25 at low $\xj$ values to 1.15 at $\xj$ values approaching unity. For per-pair normalized $\xj$ distributions, the magnitude of \cxj\ is consistent with unity and it is not applied on per-pair normalized $\xj$ distributions. Similar scaling was also applied in other analyses using 5.44 TeV \xexe data \cite{HION-2018-17,CMS-HIN-18-004}.
 
The unfolded $\ptonepttwo$ distributions are also projected onto the $\ptlead$ and $\ptsub$ axes to construct the pair nuclear-modification factors for dijets as a function of the leading jet \pt,
$$
R_{\mathrm{AA}}^{\mathrm{pair}}(\ptlead) = \frac{\frac{1}{\TAAavr N_\mathrm{evt}^{\mathrm{AA}}} \int_{0.32\times\ptlead}^{\ptlead} \frac{\dd^2N^{\mathrm{pair}}(\mathrm{AA})}{\dd\ptlead\dd\ptsub}\dd\ptsub }{\frac{1}{\mathcal{L}_{pp}} \int_{0.32\times\ptlead}^{\ptlead} \frac{\dd^2N^{\mathrm{pair}}(pp)}{\dd\ptlead\dd\ptsub} \dd\ptsub },
$$
and as a function of subleading jet \pt,
\begin{equation}
R_{\mathrm{AA}}^{\mathrm{pair}}(\ptsub) = \frac{\frac{1}{\TAAavr N_\mathrm{evt}^{\mathrm{AA}}} \int_{\ptsub}^{\ptsub/0.32} \frac{\dd^2N^{\mathrm{pair}}(\mathrm{AA})}{\dd\ptlead\dd\ptsub}\dd\ptlead }{\frac{1}{\mathcal{L}_{pp}} \int_{\ptsub}^{\ptsub/0.32} \frac{\dd^2N^{\mathrm{pair}}(pp)}{\dd\ptlead\dd\ptsub} \dd\ptlead}.
\label{eq:F:raa_sub}
\end{equation}
Here $\mathcal{L}_{pp}$ is the integrated luminosity of $pp$ collisions and the boundaries in the integrals are given by the measured minimum value of $\xj$.
 
To evaluate the differences between the dijet quenching in \xexe and \pbpb collisions the ratio of pair nuclear-modification factors for the leading jet is defined as
\begin{eqnarray}
\nonumber
\rhoab(\ptlead) &=& \frac{\RAA^{\mathrm{pair}}(\ptlead)|_{\mathrm{\xexe}}}{ \RAA^{\mathrm{pair}}(\ptlead)|_{\mathrm{\pbpb}}} = \\
&=&
\nonumber
\frac{\frac{1}{\TAAXeXe N_\mathrm{evt}^{\mathrm{\xexe}}} \int_{0.32\times\ptlead}^{\ptlead} \frac{\dd^2N^{\mathrm{pair}}(\mathrm{\xexe},5.44~\mathrm{\TeV})}{\dd\ptlead\dd\ptsub}\dd\ptsub}{\mathcal{C}(\ptlead) \times\frac{1}{\TAAPbPb N_\mathrm{evt}^{\mathrm{\pbpb}}} \int_{0.32\times\ptlead}^{\ptlead} \frac{\dd^2N^{\mathrm{pair}}(\mathrm{\pbpb},5.02~\mathrm{\TeV})}{\dd\ptlead\dd\ptsub}\dd\ptsub }.
\end{eqnarray}
As there is no reference \pp\ data at 5.44 \TeV, the factor $\mathcal{C}(\ptlead)$ is introduced to account for the difference between the center-of-mass energies of 5.44~\TeV\ and 5.02~\TeV\ collision data. It is evaluated using \PYTHIA[8] Monte Carlo simulations as
\begin{equation}
\mathcal{C}(\ptlead) = \frac{
\int_{0.32\times\ptlead}^{\ptlead} \frac{\dd^2N^{\mathrm{pair}}_{\mathrm{\PYTHIA8}}(pp,5.44~\mathrm{\TeV})}{\dd\ptlead\dd\ptsub}\dd\ptsub }{\int_{0.32\times\ptlead}^{\ptlead} \frac{\dd^2N^{\mathrm{pair}}_{\mathrm{\PYTHIA8}}(pp,5.02~\mathrm{\TeV})}{\dd\ptlead\dd\ptsub}\dd\ptsub }.
\label{eq:F:C_lead}
\end{equation}
 
Analogously, the ratio of pair nuclear-modification factors for subleading jets, $\rhoab(\ptsub)$, and the factor $\mathcal{C}(\ptsub)$ can be defined using Equation~(\ref{eq:F:raa_sub}).
The magnitude of the $\mathcal{C}$ factor ranges from 1.12 at 32~\GeV\ to 1.31 at 398~\GeV\ for both leading and sub-leading jets.
The ratios  $\rhoab(\ptlead)$ and $\rhoab(\ptsub)$ allow the differences between the jet quenching in \xexe and \pbpb collisions to be directly quantified.

\section{Systematic uncertainties}
 
Systematic uncertainties arise from uncertainties in the JES, JER, background subtraction procedures, \TAAavr\ values, the unfolding weight selection, minimum $\ptsub$ lower boundary, and from the unfolding procedure performance observed in the MC sample. For each source of systematic uncertainty, except for the uncertainty in \TAAavr\ and unfolding non-closure, the entire analysis is repeated by varying the response matrix according to the systematic uncertainties. The difference between the final distributions of the baseline measurement and the measurement with varied values is used as the estimate of the systematic uncertainty. For the \rhoab, the JES and JER systematic uncertainties are correlated between Xe+Xe and Pb+Pb, while all the other uncertainties are taken as uncorrelated.
 
The systematic uncertainty in the JES has four components. The first, centrality-independent component, is determined from in situ studies of the calorimeter response to jets reconstructed with the procedure used in 13~\TeV \pp\ collisions~\cite{PERF-2011-03,PERF-2016-04}. The second component accounts for the relative energy-scale difference between the jet reconstruction procedures used in this analysis and those used for 13~\TeV \pp\ collisions~\cite{ATLAS-CONF-2015-016}. The third component accounts for possible mis-modeling of the relative abundances of quark and gluon jets and the calorimeter response to them in the MC simulation. This is assessed by comparing quark and gluon jets generated with \PYTHIA and with \HERWIGpp. The fourth component is centrality dependent and accounts for any incomplete knowledge of the calorimetric response to quenched jets through modifications of parton showering~\cite{ATLAS-CONF-2015-016,HION-2013-06}. The total size of systematic uncertainty from the JES is approximately $1-8$\% with the maximum around $\xj \approx 0.32$ in $0-10\%$ central collisions. It decreases with $\xj$ except for the last centrality interval where it increases, reaching a value of about 2\%. For the \rhoab, the largest systematic uncertainty from the JES is 5\%  and it exhibits only a weak \pt dependence.
 
The systematic uncertainty in the JER has two components. The first component is evaluated using an in situ technique for 13~\TeV\ \pp\ data that involves studies of dijet energy balance~\cite{PERF-2014-02,PERF-2011-04}. The second component accounts for differences between the tower-based jet reconstruction and the jet reconstruction used in the analyses of 13~\TeV\ \pp\ data and the differences between the calibration procedures. Both of these uncertainties are applied via the smearing factor that is used to include an additional contribution to the resolution of the reconstructed $\pt$ in the MC sample by the Gaussian smearing procedure. This modified reconstructed $\pt$\ enters the response matrices that are used to derive the alternative result. The uncertainty from the JER  is approximately 10\% for $\xj \approx 0.32$ in 0--10\% central collisions and decreases with $\xj$ except for the last two intervals in $\xj$ where it increases to about 5\%. For $\rhoab$, the largest systematic uncertainty from the JER is 2\% and exhibits only a weak \pt dependence. The smaller JER uncertainty in \rhoab\ compared with the JES is due to a correlation with the Pb+Pb uncertainties.
 
The systematic uncertainty arising from the removal of the combinatoric jet background procedure has two components. The first component is connected with the determination of the yield of combinatoric background jets and is determined using an alternative sideband of $1.1 < \Delta\phi < 1.5$ following the procedure described in Ref.~\cite{HION-2012-11}. The second contribution is associated with the determination of the efficiency correction and is determined from the difference between the analysis performed with and without
the efficiency correction. These uncertainties do not dominate the overall systematic uncertainty. The largest values of the uncertainty due to the determination of combinatoric jet background and the uncertainty due to the efficiency correction are about 1\% and 2\%, respectively, for both the $\xj$ and \rhoab\ distributions.
 
The uncertainty in $\TAAavr$ arises from geometric modeling uncertainties (e.g. nucleon--nucleon inelastic cross-section, Woods--Saxon parameterization of the nucleon distribution~\cite{Miller:2007ri,dEnterria:2020dwq}) and the uncertainty of the fraction of selected inelastic \xexe collisions. This uncertainty only affects the overall normalization and is independent of the dijet kinematics. The values of the uncertainties in \TAAavr\ for \xexe\ are shown in Table~\ref{tab:FCalEtXeXe} and Table~\ref{tab:FCalEtXeXeCompare}. This uncertainty is uncorrelated between \xexe\ and \pbpb~\cite{HION-2019-02}.
 
There are two sources of systematic uncertainty connected with the unfolding procedure. The first one arises from the imprecision in the determination of the initial distributions used in the iterative procedure for the underlying generator-level distribution in the unfolding procedure. This uncertainty is calculated from the difference between the unfolded distributions constructed using the nominal reweighted prior and the prior without applying the reweighting. The second source derives from the sensitivity of the unfolding procedure to the jet selection choice of the minimum jet $\pt$. This is estimated by changing the minimum jet $\pt$ from 32~\GeV\ to 25~\GeV. To perform a check on the performance of the full analysis procedure a closure test is performed with the MC sample by evaluating the differences between the final unfolded distributions and the generator-level distributions of the MC sample. The difference from unity in the closure test is included as the additional source of systematic uncertainty. These uncertainties are the largest at low $\xj$, where the uncertainty due to the prior sensitivity reaches 10\% for $100 < \ptlead < 126$~\GeV\ in 0--10\% central collisions. For larger $\xj$ and \pt they decrease to 1--3\%.

The uncertainty in the $\mathcal{C}(\pt)$ factors defined in Section~\ref{sec:analysis} is estimated as the difference between the $\mathcal{C}(\pt)$ factors evaluated using \PYTHIA[8] and \HERWIGpp MC samples. The magnitude of the uncertainty in the $C$ factor stays below 2\% over the full \pt range of the \rhoab\ distribution.
The uncertainty in the \cxj\ factor is evaluated in the same way as the uncertainty in the $\mathcal{C}(\pt)$ factors. Its magnitude stays below 2\%, and it is applied as the additional uncertainty in the \pbpb $\xj$ distributions.
 
\begin{figure}
\centering
\centerline{
\includegraphics[width=0.33\linewidth]{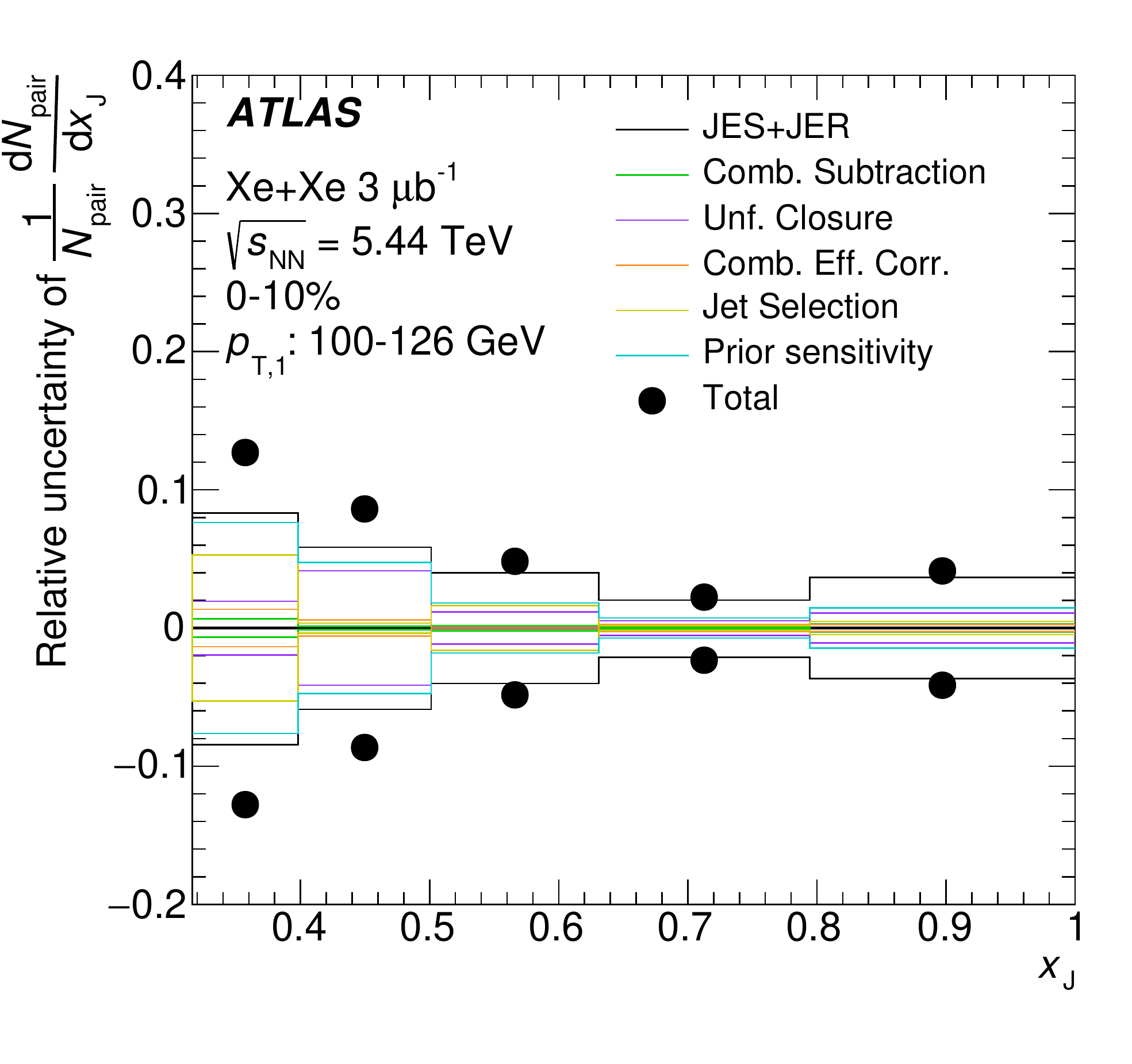}
\includegraphics[width=0.33\linewidth]{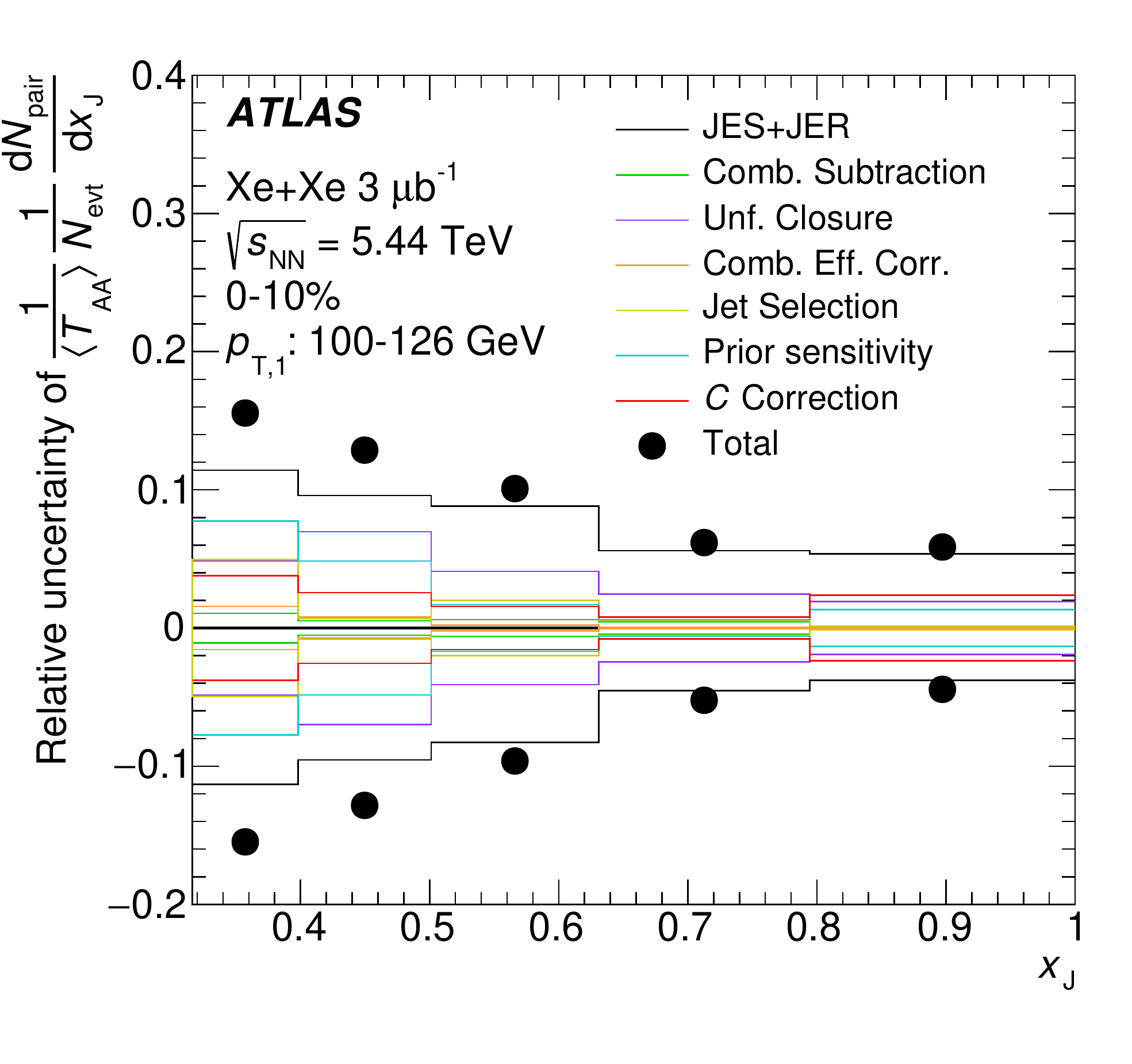}
\includegraphics[width=0.33\linewidth]{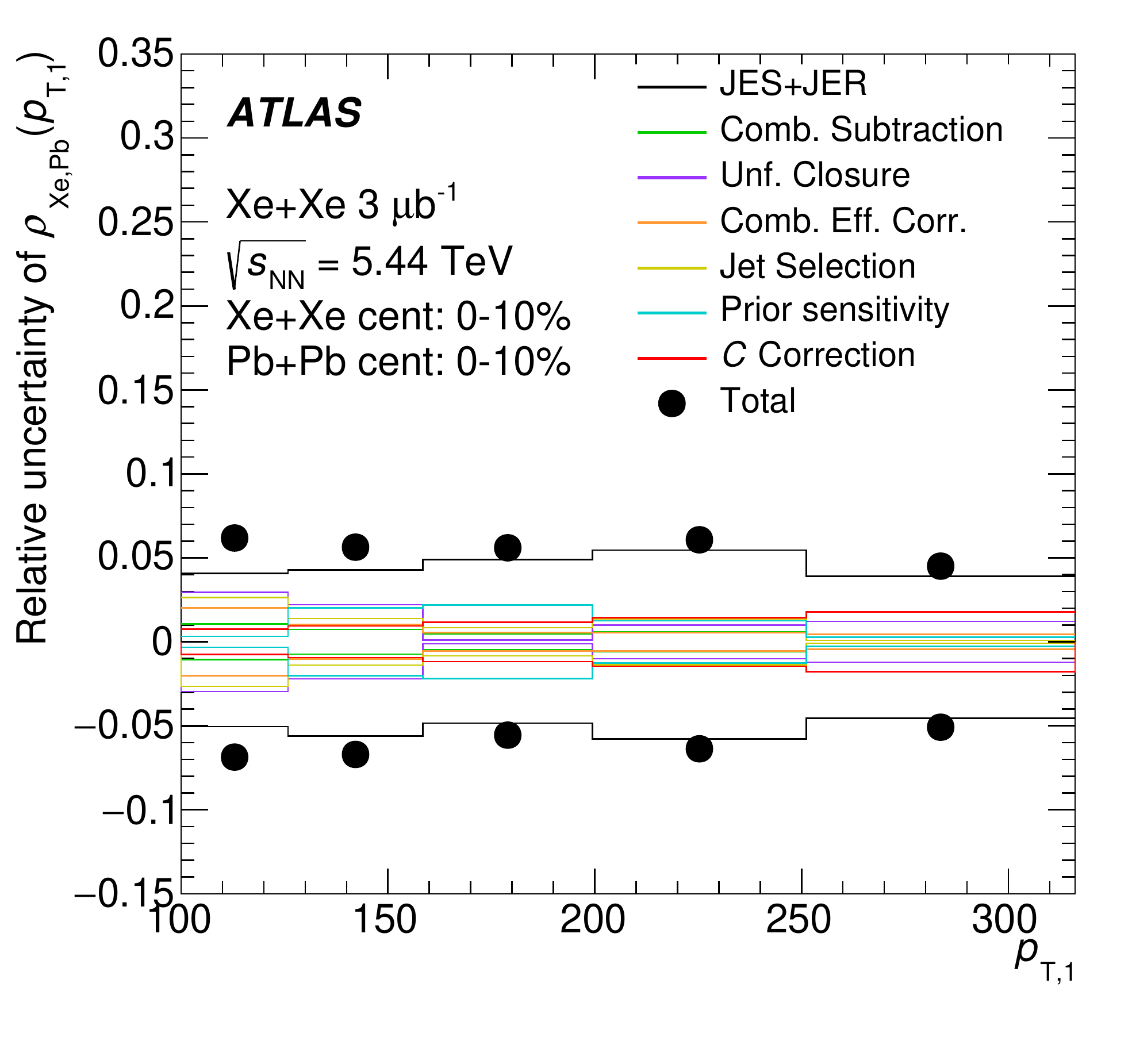}
}
\caption{The relative magnitude of systematic uncertainties for (left) the per-pair normalized $\xj$ distribution, (middle) the absolutely normalized $\xj$ distribution, and (right) the $\rhoab(\ptlead)$, all in the 0--10\% centrality interval.}
\label{fig:sys}
\end{figure}

The magnitude of systematic uncertainties for per-pair normalized dijet momentum-imbalance, absolutely normalized dijet momentum-imbalance, and for the $\rhoab(\ptlead)$ distributions, in the most central collisions, is shown in Figure~\ref{fig:sys}.
 
\section{Results}
\begin{figure}
\centering
\centerline{
\includegraphics[width=0.35\linewidth]{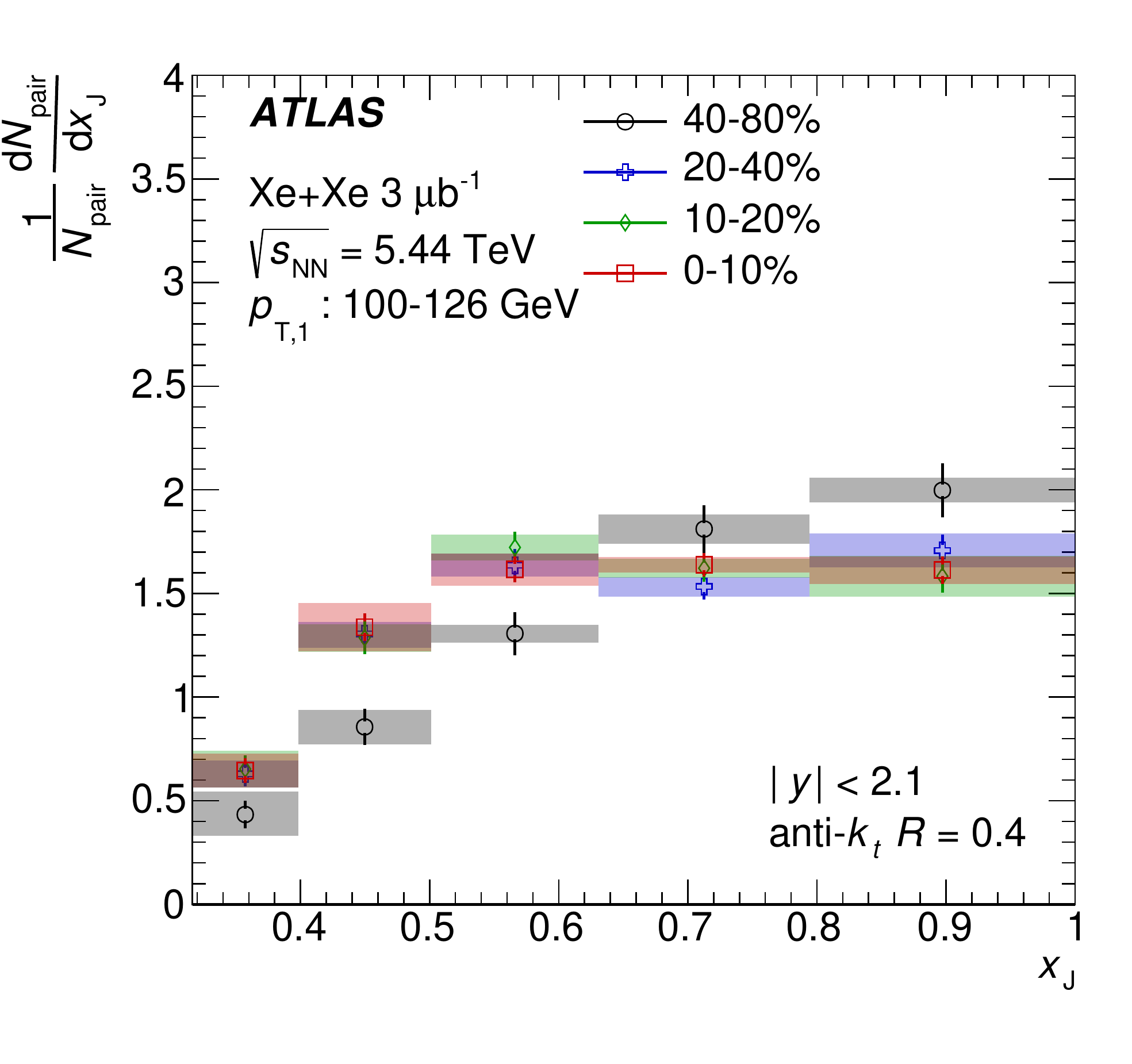}
\includegraphics[width=0.35\linewidth]{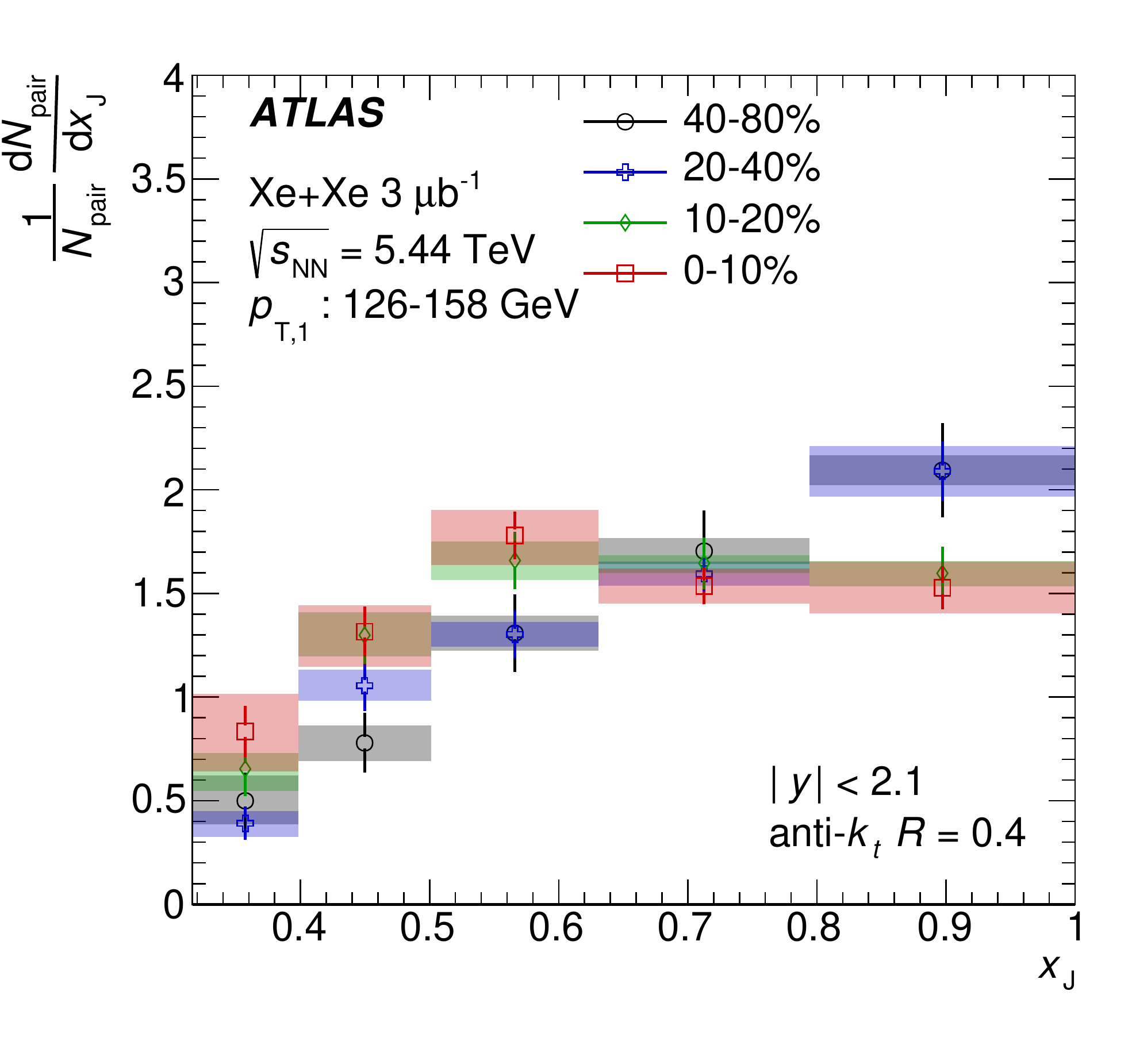}
\includegraphics[width=0.35\linewidth]{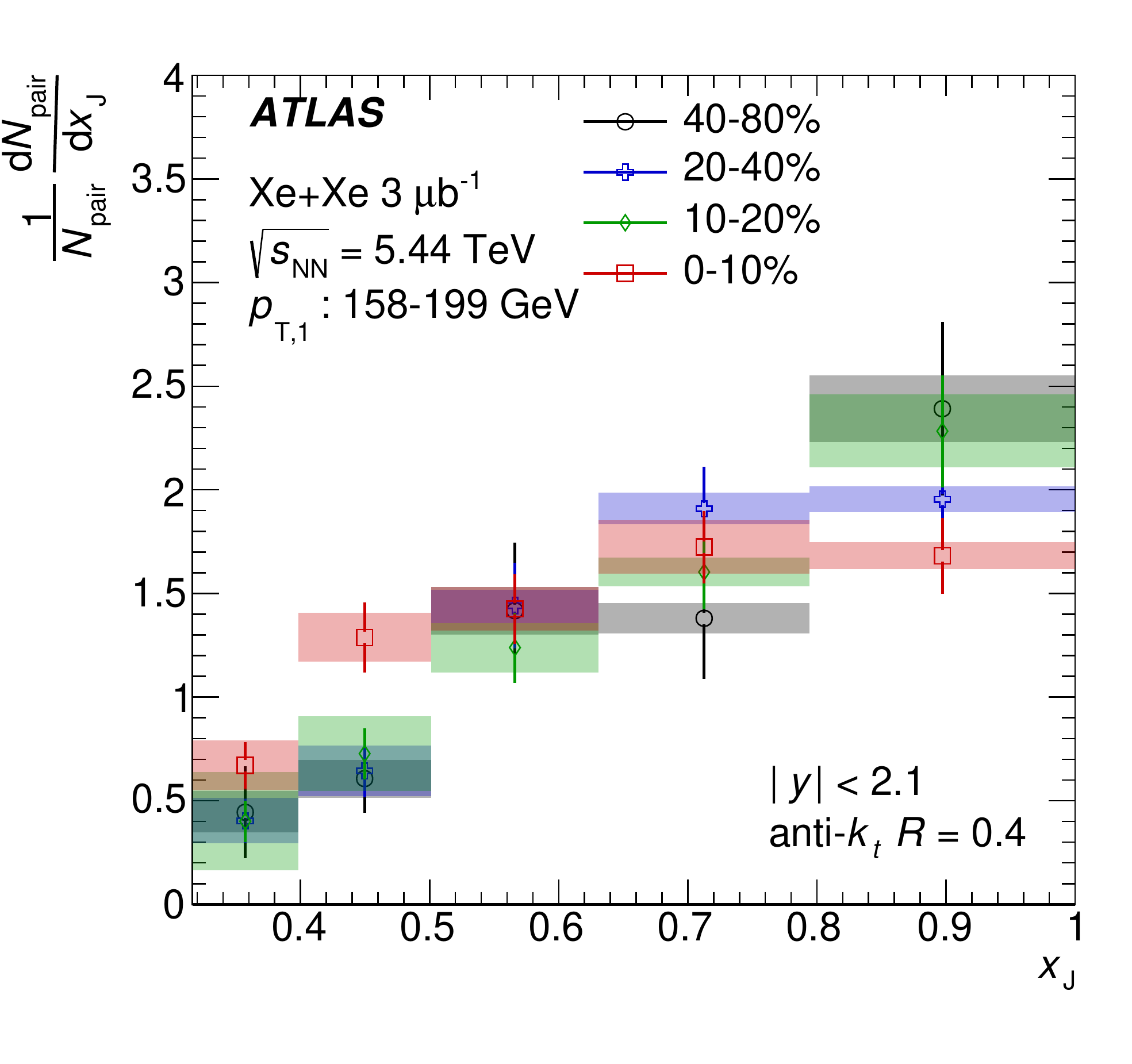}}
\caption{Per-pair normalized $\xj$ distribution evaluated in four centrality intervals and three $\pt$ intervals of leading jet \pt: (left) $100 < \ptlead < 126$~\GeV, (middle) $126 < \ptlead < 158$~\GeV, and (right) $158 < \ptlead < 199$~\GeV. Statistical and systematic uncertainties are represented by error bars and boxes, respectively.}
\label{fig:xj-pair}
\end{figure}
 
\begin{figure}
\centering
\centerline{
\includegraphics[width=0.35\linewidth]{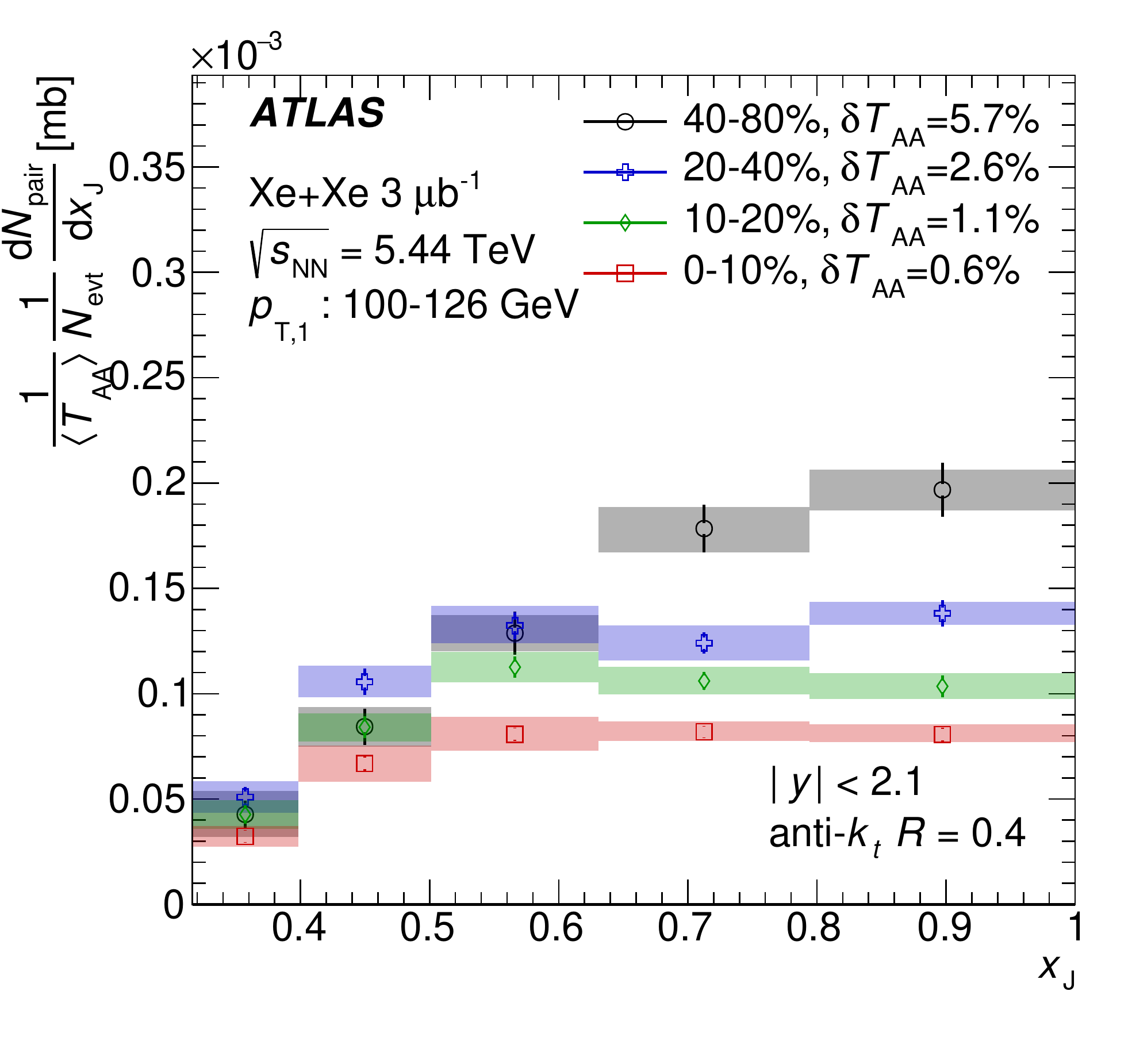}
\includegraphics[width=0.35\linewidth]{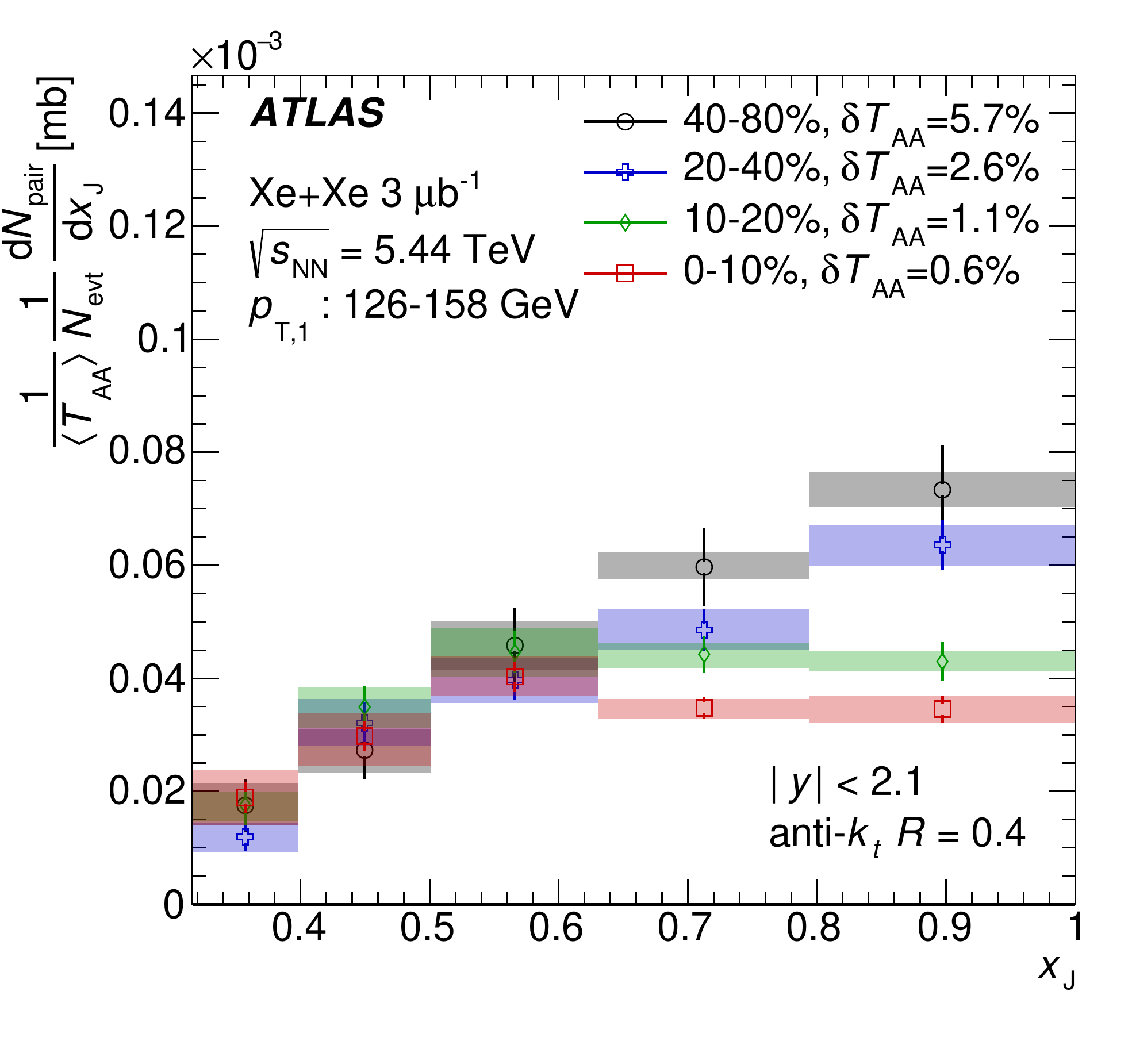}
\includegraphics[width=0.35\linewidth]{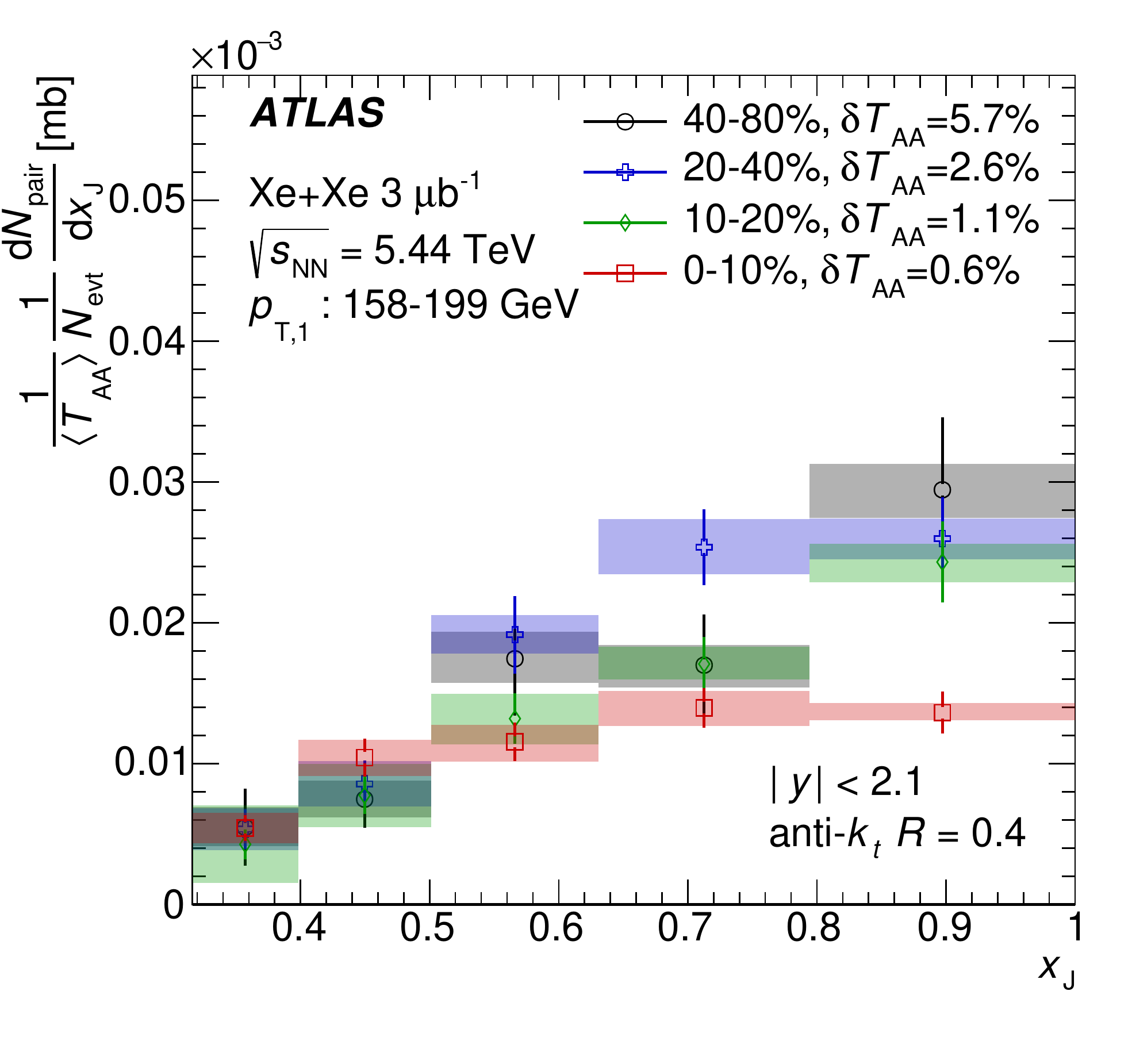}}
\caption{Absolutely normalized $\xj$ distribution evaluated in four centrality intervals and three $\pt$ intervals of leading jet \pt: (left) $100 < \ptlead < 126$~\GeV, (middle) $126 < \ptlead < 158$~\GeV, and (right) $158 < \ptlead < 199$~\GeV.  Statistical and systematic uncertainties are represented by error bars and boxes, respectively. The $\delta T_\text{AA}$ in the legend represents the relative uncertainty on $\TAAavr$.}
\label{fig:xj-abs}
\end{figure}
 
Figure~\ref{fig:xj-pair} shows the per-pair normalized distribution of $\xj$ evaluated in four centrality intervals (0--10\%, 10--20\%, 20--40\%, and 40--80\%) and
three $\pt$ intervals of the leading jet \pt ($100 < \ptlead < 126$~\GeV, $126 < \ptlead < 158$~\GeV, and $158 < \ptlead < 199$~\GeV). A substantial difference
between the shape of $\xj$ distribution in the most central collisions (0--10\%) and the most peripheral collisions (40--80\%) is seen. In peripheral collisions, the
most frequent configurations are balanced dijets, while in central collisions the rate of imbalanced dijets is the same or higher than the rate of balanced dijets.
These features are also observed in \pbpb collisions~\cite{HION-2019-02}.
Narrowing of the $\xj$ distribution with increasing $\ptlead$, previously measured in
\pbpb\ collisions, is not that pronounced in the measured \pt interval in \xexe collisions.
The peak structure at $\xj=0.6$, previously measured in 0--10\% \pbpb collisions, is not present in 0--10\% \xexe collisions. This
may be connected with a smaller overlapping region of colliding nuclei in \xexe compared with \pbpb collisions. The evolution between the central and
peripheral \xexe collisions is not as pronounced as in \pbpb collisions. 
The absence of a
clearly visible evolution is connected with a worse statistical precision of the \xexe measurement compared with the \pbpb measurement.
 
Figure~\ref{fig:xj-abs} shows the absolutely normalized distribution of $\xj$ evaluated for the same centrality and $\ptlead$ selection as in Figure~\ref{fig:xj-pair}. It shows that the relative enhancement of imbalanced dijet topologies seen in Figure~\ref{fig:xj-pair} is due to the depletion in the absolute yield of balanced dijets -- an observation valid also in the~\pbpb measurement. The results in Figure~\ref{fig:xj-abs} show a clear centrality evolution where the suppression of the balanced dijet yield gradually decreases from central to peripheral collisions.
 
\begin{figure}
\centering
\centerline{
\includegraphics[width=0.33\linewidth]{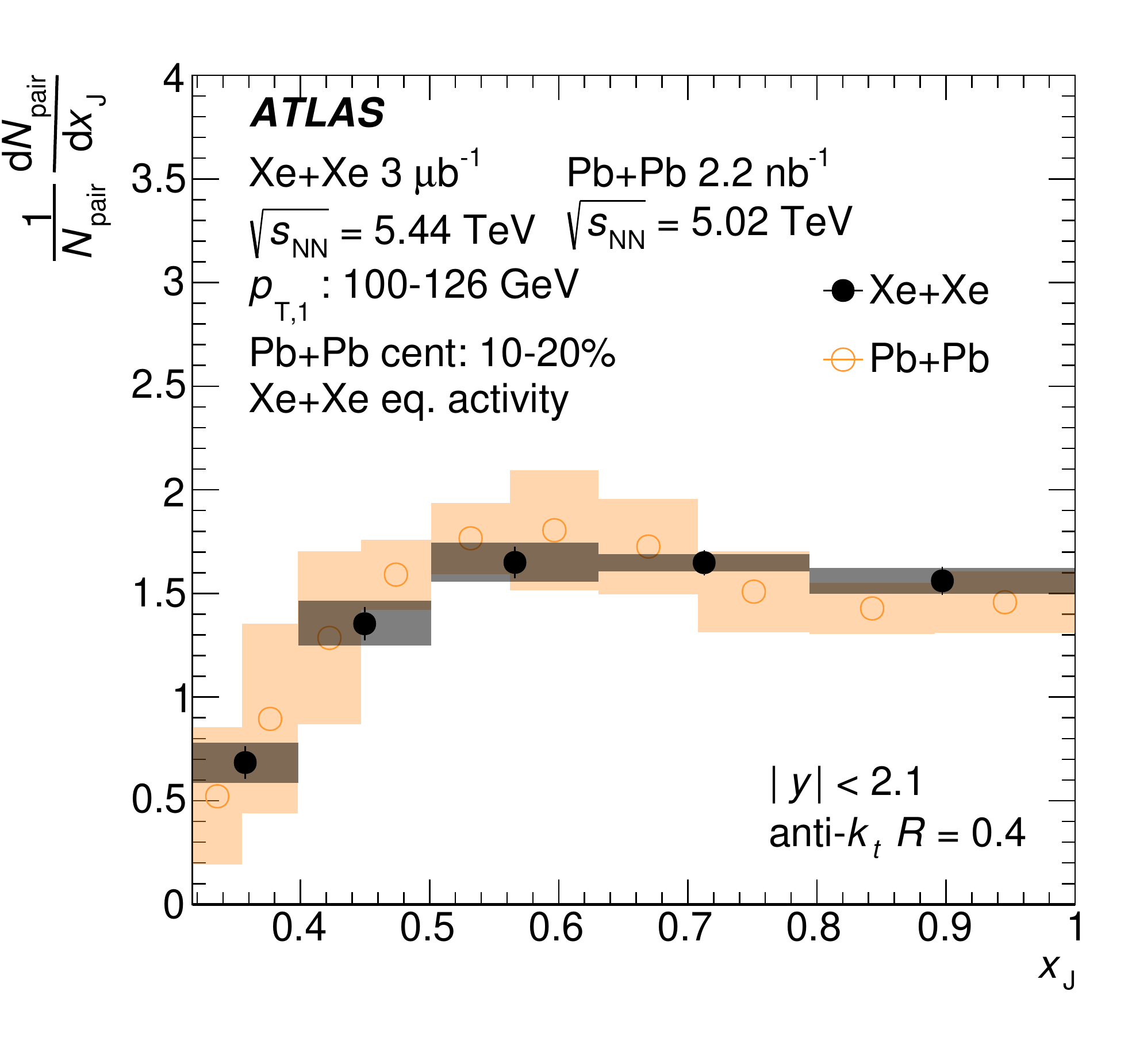}
\includegraphics[width=0.33\linewidth]{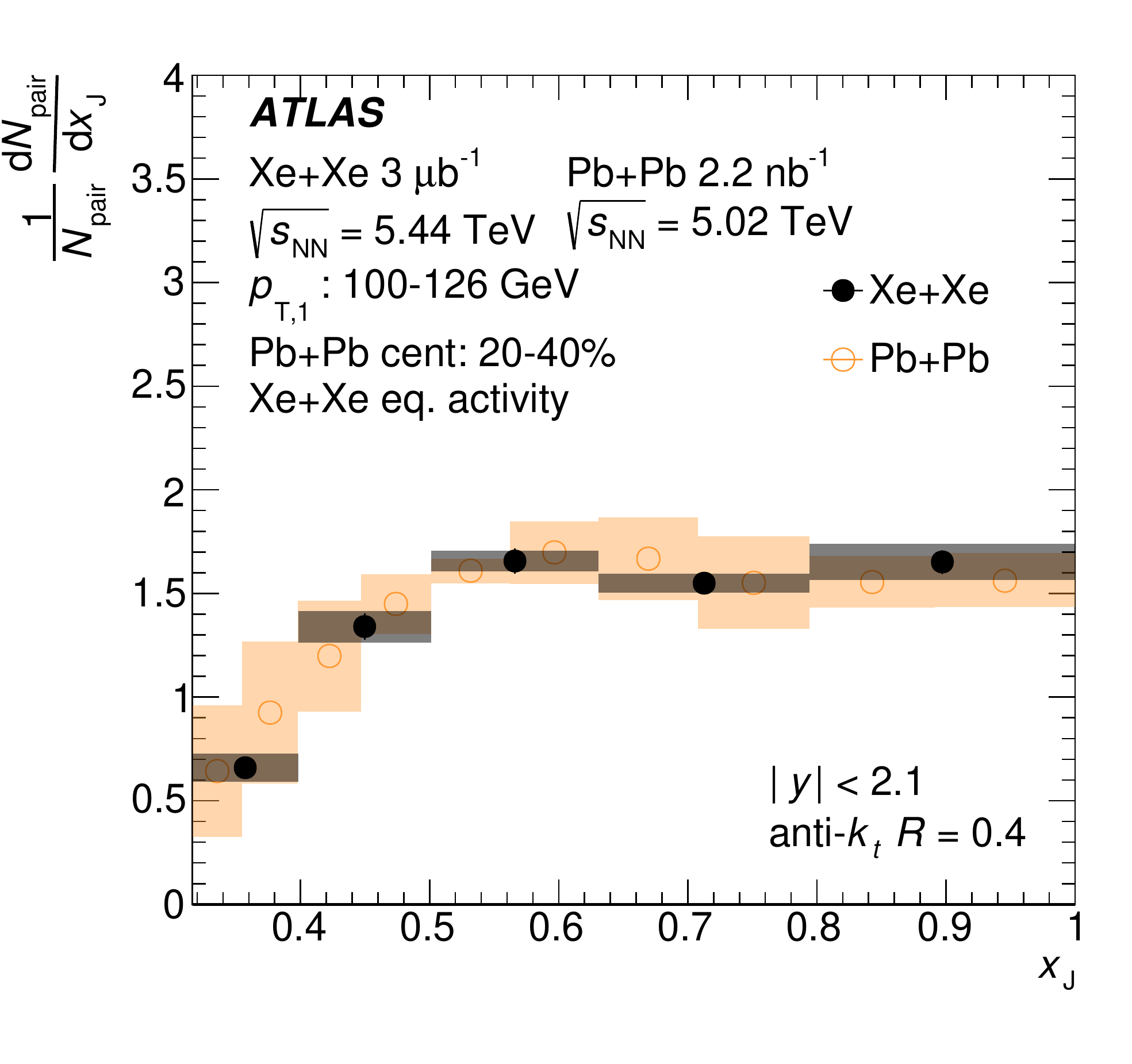}
\includegraphics[width=0.33\linewidth]{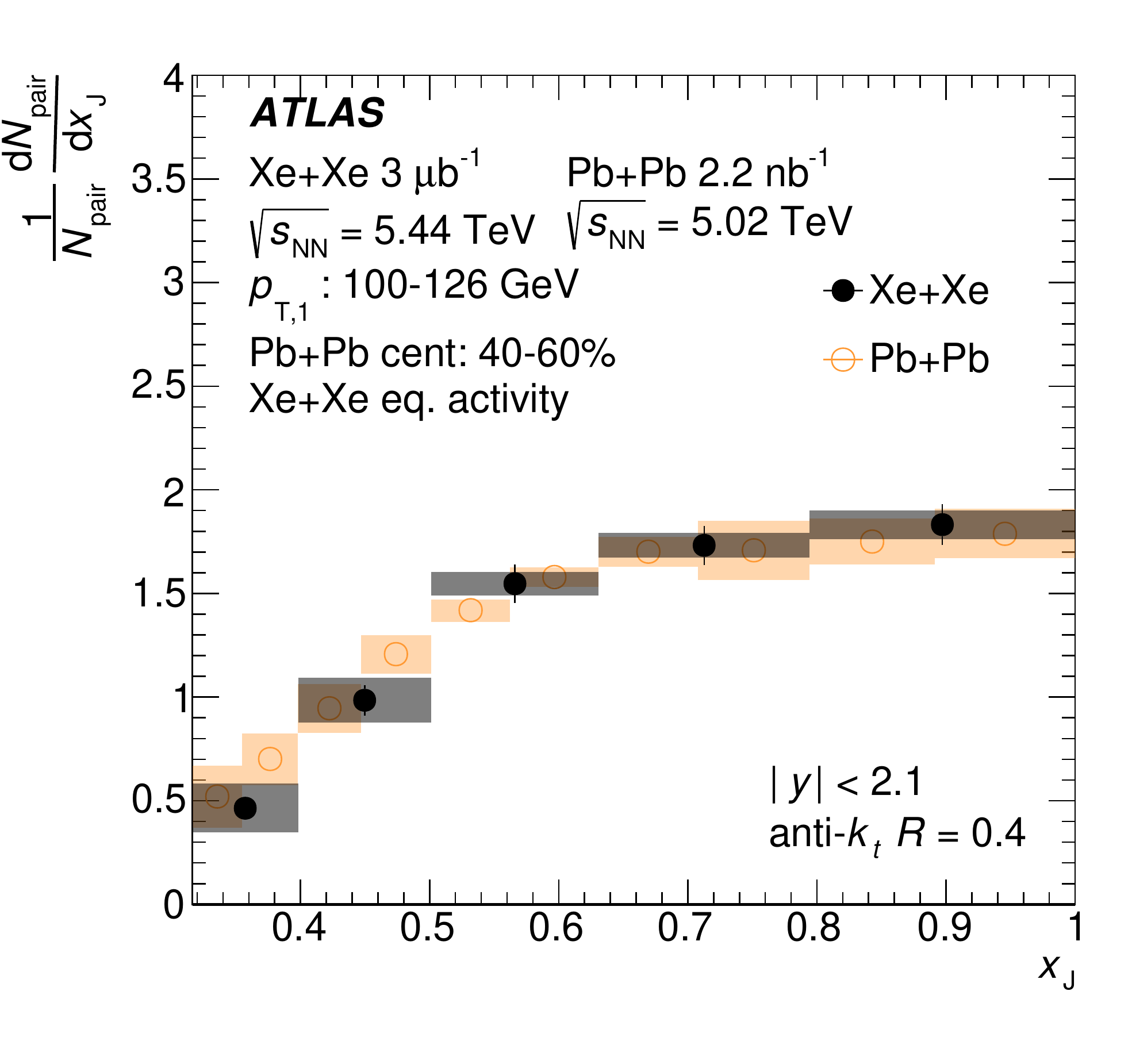}}
\caption{Comparison of \xexe (filled points) and \pbpb (open points) per-pair normalized $\xj$ distributions in $100~<~\ptlead~<~126$~\GeV\ and \centrality{10}{20}, \centrality{20}{40}, and \centrality{40}{60} \pbpb\ centrality intervals and in the corresponding \xexe\ \sumETFCal\ intervals. Statistical and systematic uncertainties are represented by error bars and boxes, respectively.}
\label{fig:xj-pbpb-fcal-pair}
\end{figure}

To compare the $\xj$ distribution between \xexe and \pbpb in an different way, the $\xj$ distributions are evaluated in intervals of the same event activity, quantified by \sumETFCal. The choice of \sumETFCal intervals matches those measured in \pbpb for centrality intervals 10--20\%, 20--40\%, and 40--60\%. The corresponding centrality intervals in \xexe collisions are given in Table~\ref{tab:FCalEtXeXeCompare}. The most central \pbpb interval (0--10\%) cannot be used since the equivalent event activity is not present in \xexe collisions. The \xexe to \pbpb comparison of per-pair normalized $\xj$ distributions is presented in Figure~\ref{fig:xj-pbpb-fcal-pair} in the $100 < \ptlead < 126$~\GeV\ interval. The distributions measured within the same event activity interval are consistent between \xexe and \pbpb collisions. A similar agreement is also found in other $\ptlead$ selections. The smaller systematic uncertainties for the lower statistic \xexe collisions are connected with the coarser binning used in \xexe data, which results in smaller bin-to-bin migrations and, consequently, smaller systematic uncertainties related to the unfolding procedure.
 
\begin{figure}
\centering
\centerline{
\includegraphics[width=0.33\linewidth]{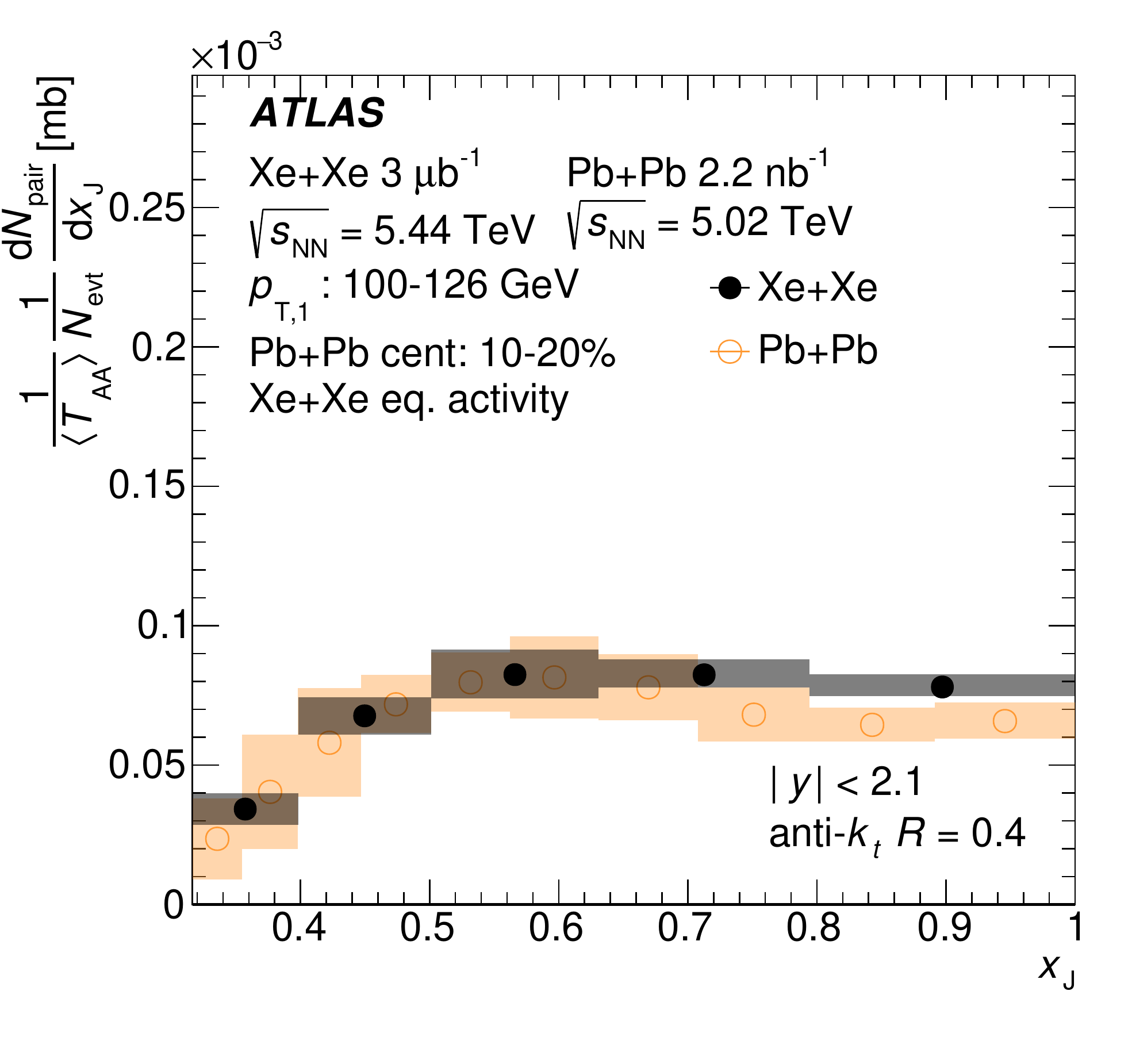}
\includegraphics[width=0.33\linewidth]{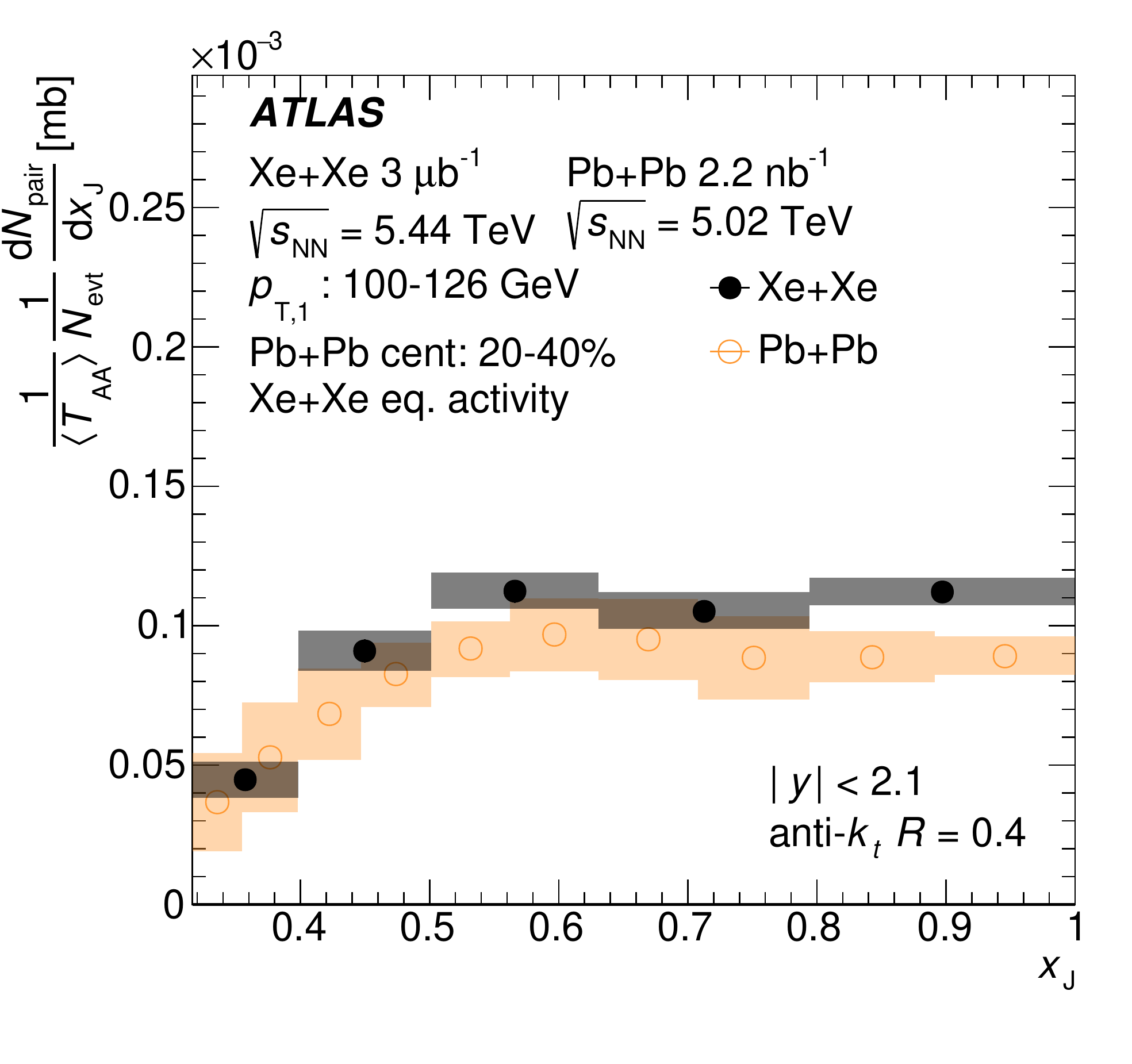}
\includegraphics[width=0.33\linewidth]{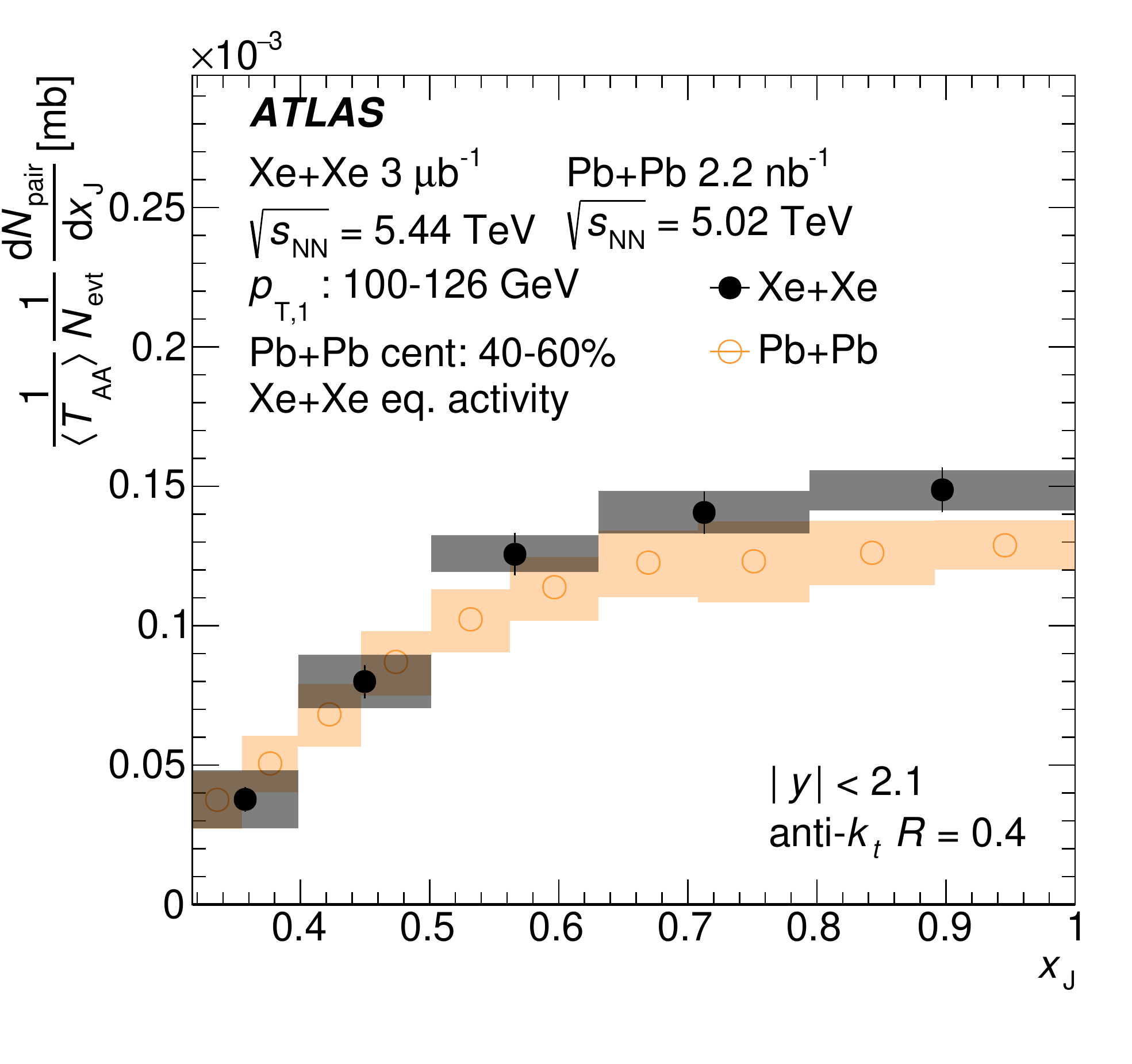}}
\centerline{
\includegraphics[width=0.33\linewidth]{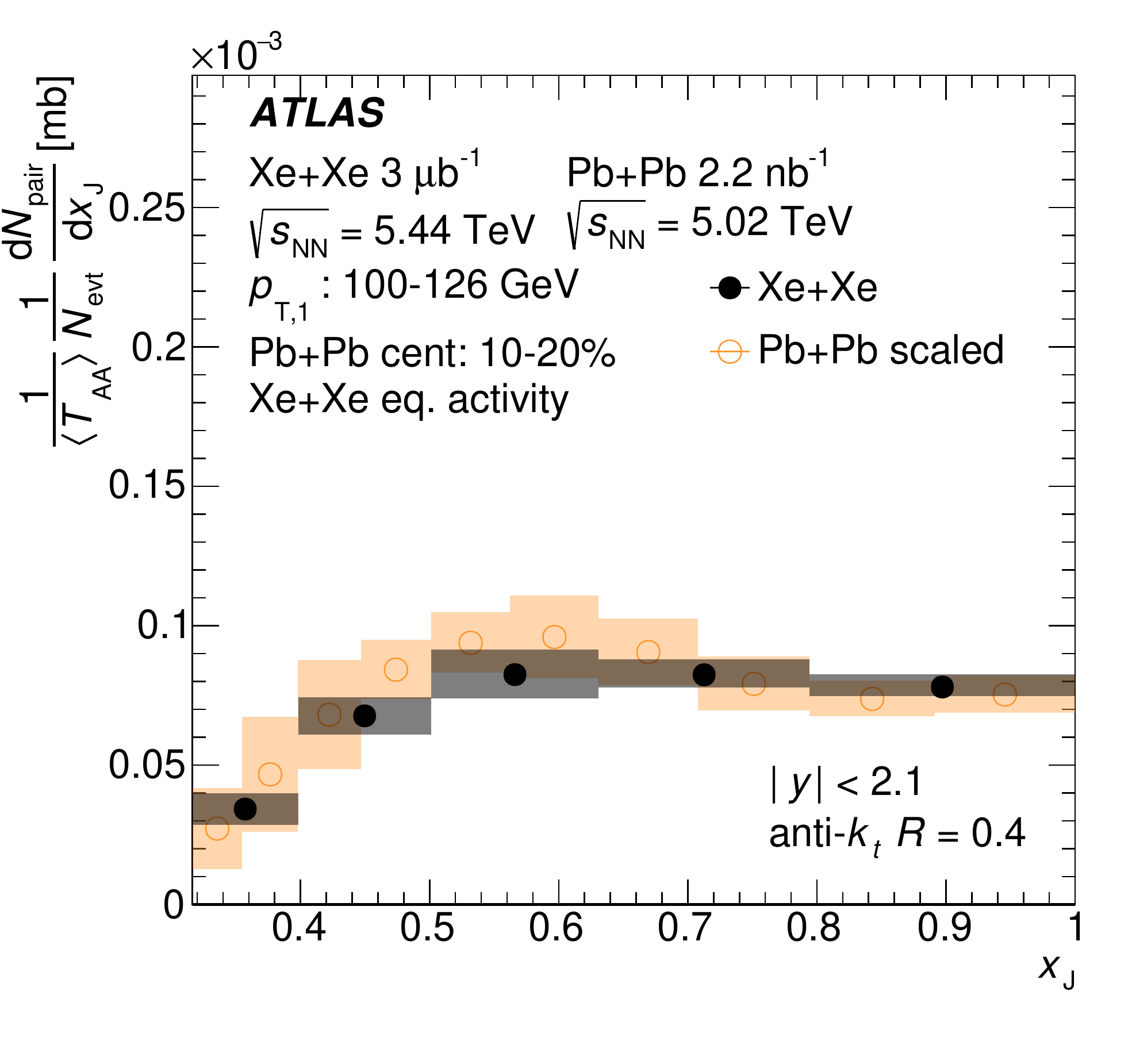}
\includegraphics[width=0.33\linewidth]{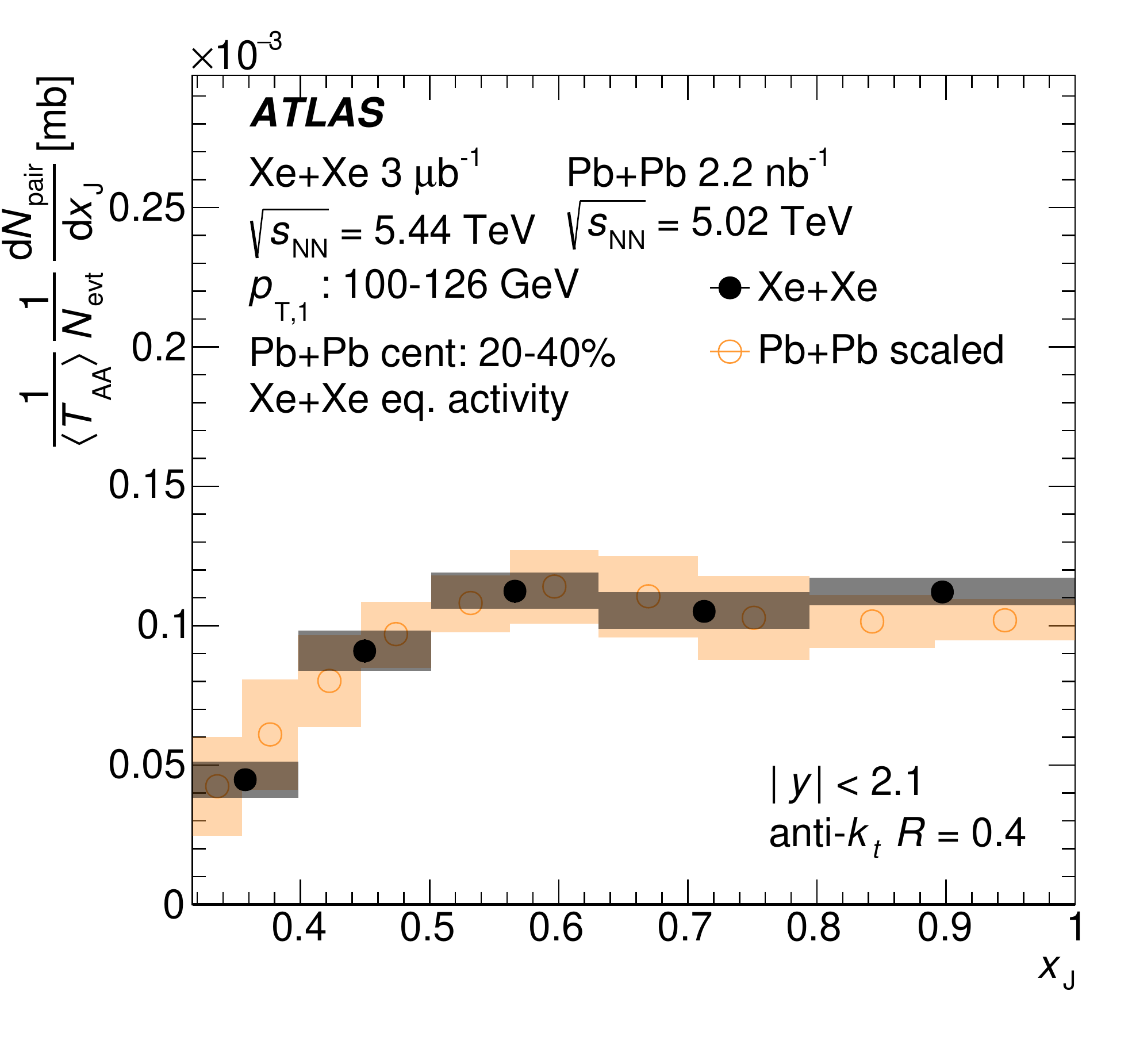}
\includegraphics[width=0.33\linewidth]{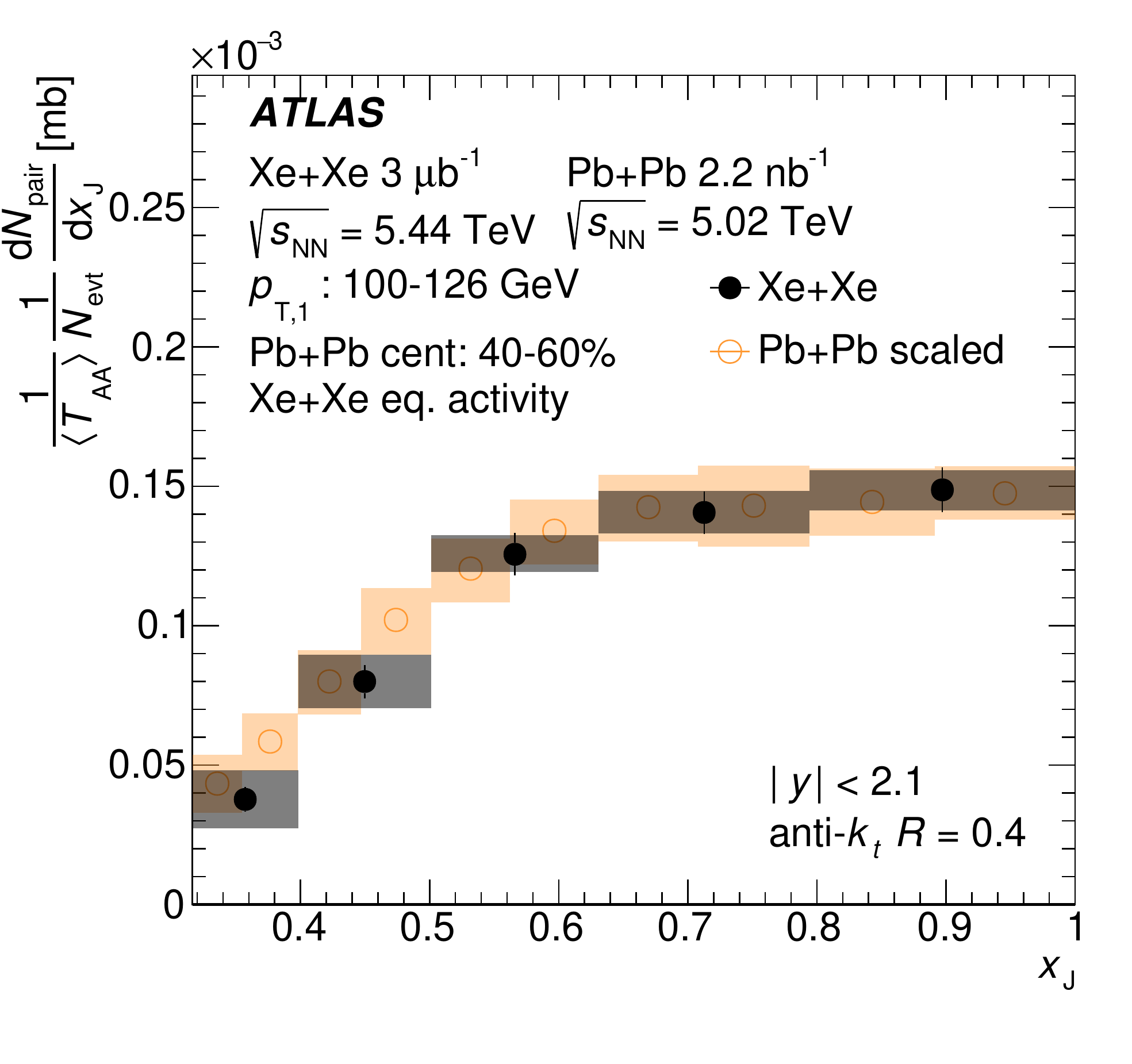}}
 
\caption{Comparison of \xexe (filled points) and \pbpb (open points) absolutely normalized $\xj$ distributions in $100~<~\ptlead~<~126$~\GeV\ and \centrality{10}{20}, \centrality{20}{40}, and \centrality{40}{60} \pbpb\ centrality intervals and in the corresponding \xexe\ \sumETFCal\ intervals. The upper plots show directly measured distributions. The lower plots show the \pbpb distributions corrected for the impact of the difference between the center-of-mass energies of the hard scattering process in \xexe and \pbpb collisions. Statistical and systematic uncertainties are represented by error bars and boxes, respectively.}
\label{fig:xj-pbpb-fcal-abs}
\end{figure}

The comparison of absolutely normalized $\xj$ distributions between \pbpb\ and \xexe\ in the same event activity intervals is presented in the upper plots of Figure~\ref{fig:xj-pbpb-fcal-abs}. A clear difference between \xexe and \pbpb distributions can be seen, with \xexe having a larger absolute yield than \pbpb.
This difference may be partially attributed to the difference between the hard process cross-sections, due to the different center-of-mass energies of the initial hard scattering, in \xexe and \pbpb collisions.
To estimate the impact of the difference between the center-of-mass energies, the absolutely
normalized $\xj$ distributions in \pbpb\ collisions are scaled by $\cxj$ defined in Equation~(\ref{eq:cxj}). The result is shown in the bottom plots of Figure~\ref{fig:xj-pbpb-fcal-abs}.
After correcting for the difference between the center-of-mass energies of the initial hard scattering, the absolutely normalized $\xj$ distributions agree between \xexe and \pbpb collisions within uncertainties. The same conclusion is also found for other $\ptlead$ intervals.
While the observed agreement could arise from canceling effects and large uncertainties, a natural explanation for this behavior is that the difference between the energies of the hard scattering process plays a significant role in the absolutely normalized $\xj$ distributions.

In the case of per-pair normalized $\xj$ distributions, the correction factor $\mathcal{C}(\xj)$ was found to be consistent with unity, which is consistent with observing an agreement of per-pair normalized $\xj$ distributions between \xexe and \pbpb collisions.  \textbf{}
 
\begin{figure}
\centering
\centerline{
\includegraphics[width=0.33\linewidth]{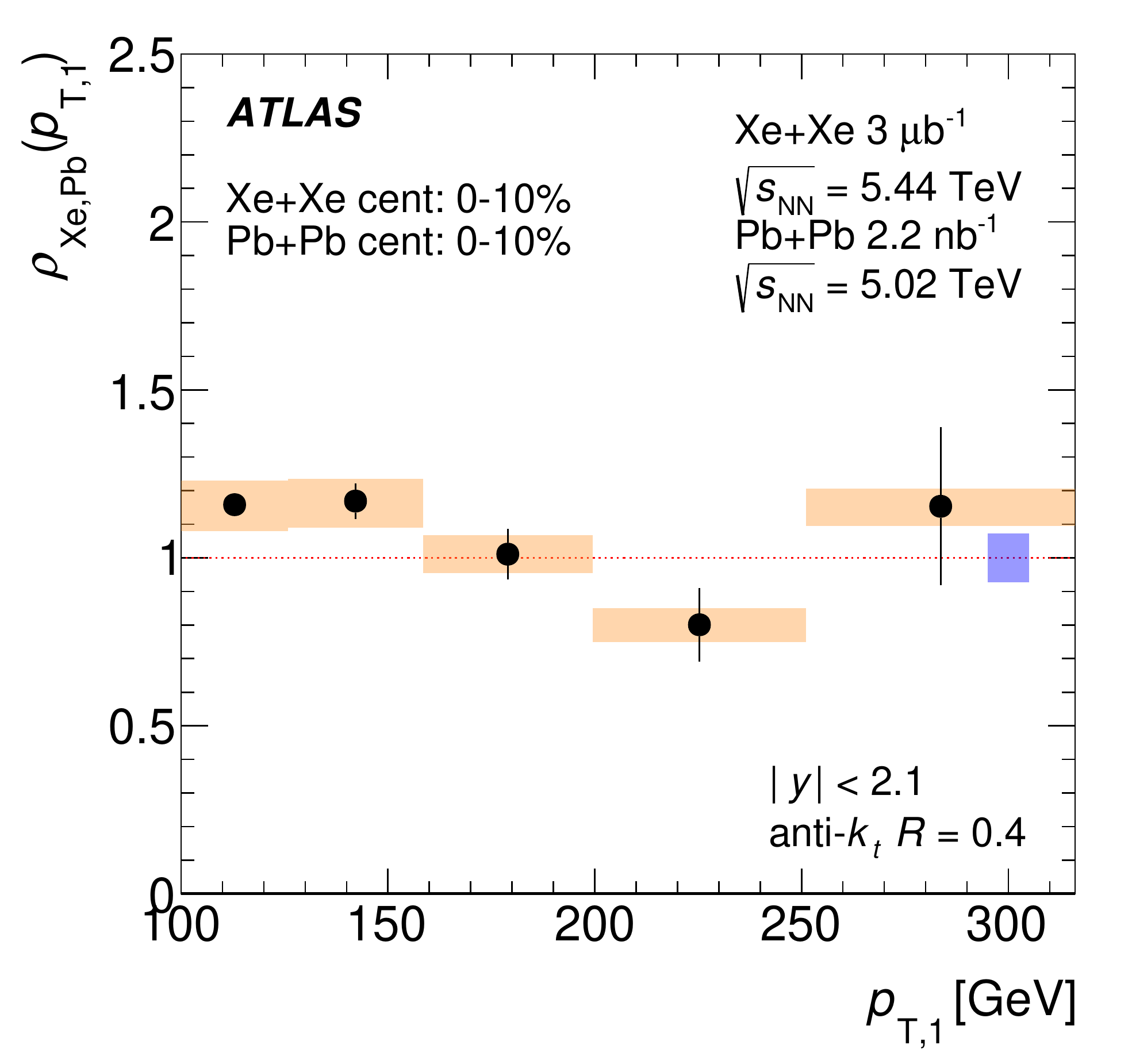}
\includegraphics[width=0.33\linewidth]{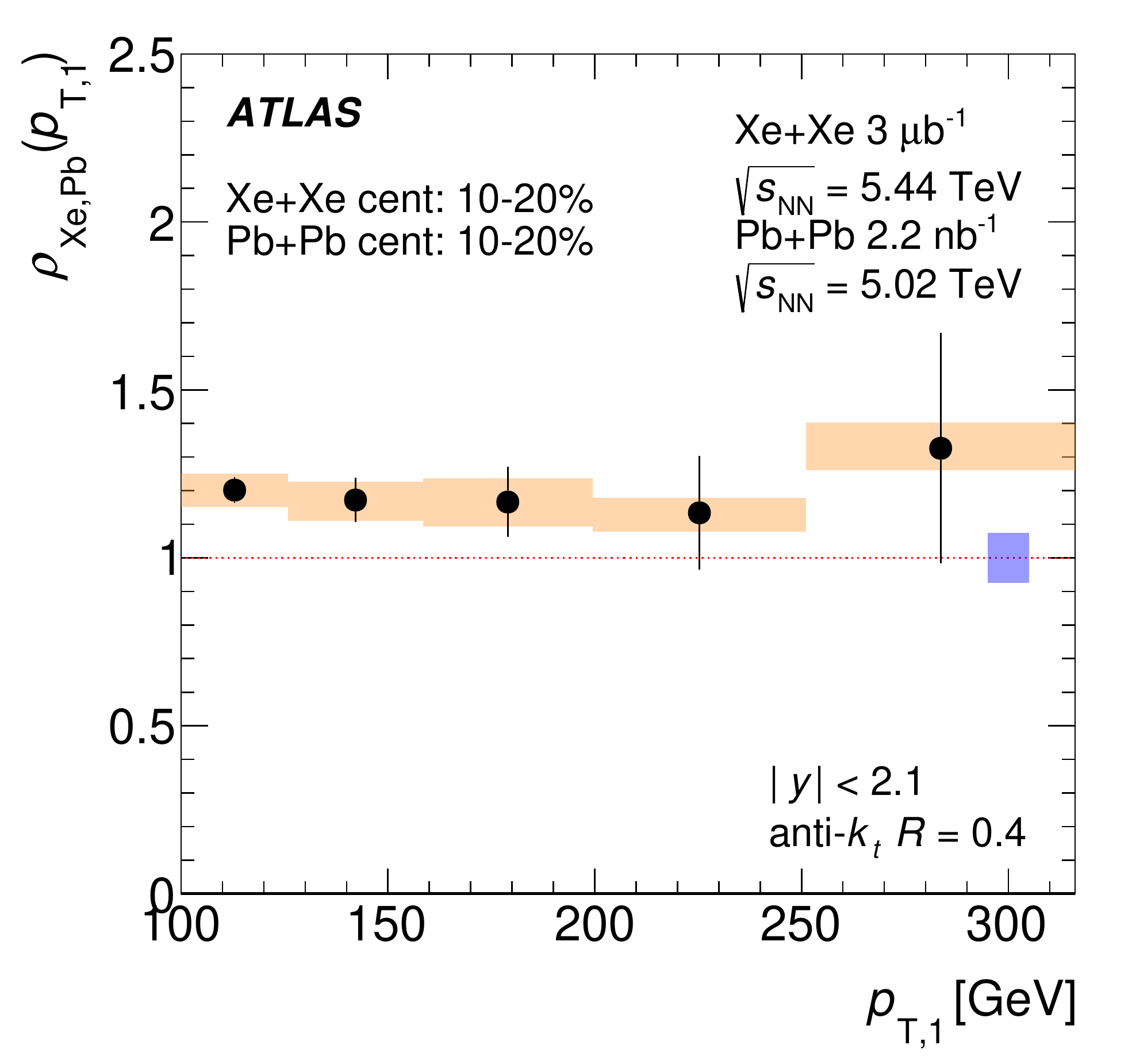}
\includegraphics[width=0.33\linewidth]{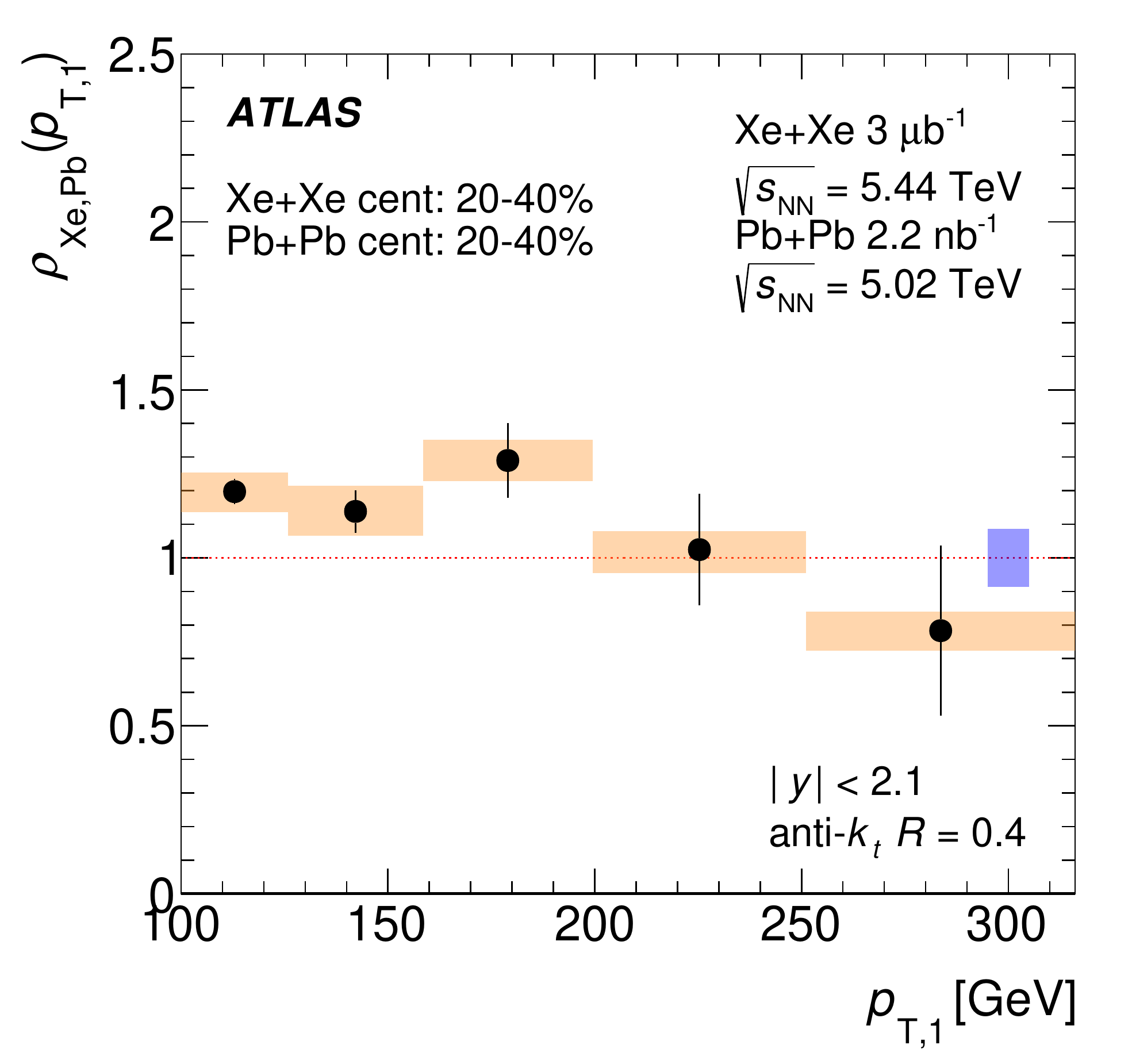}
}
\centerline{
\includegraphics[width=0.33\linewidth]{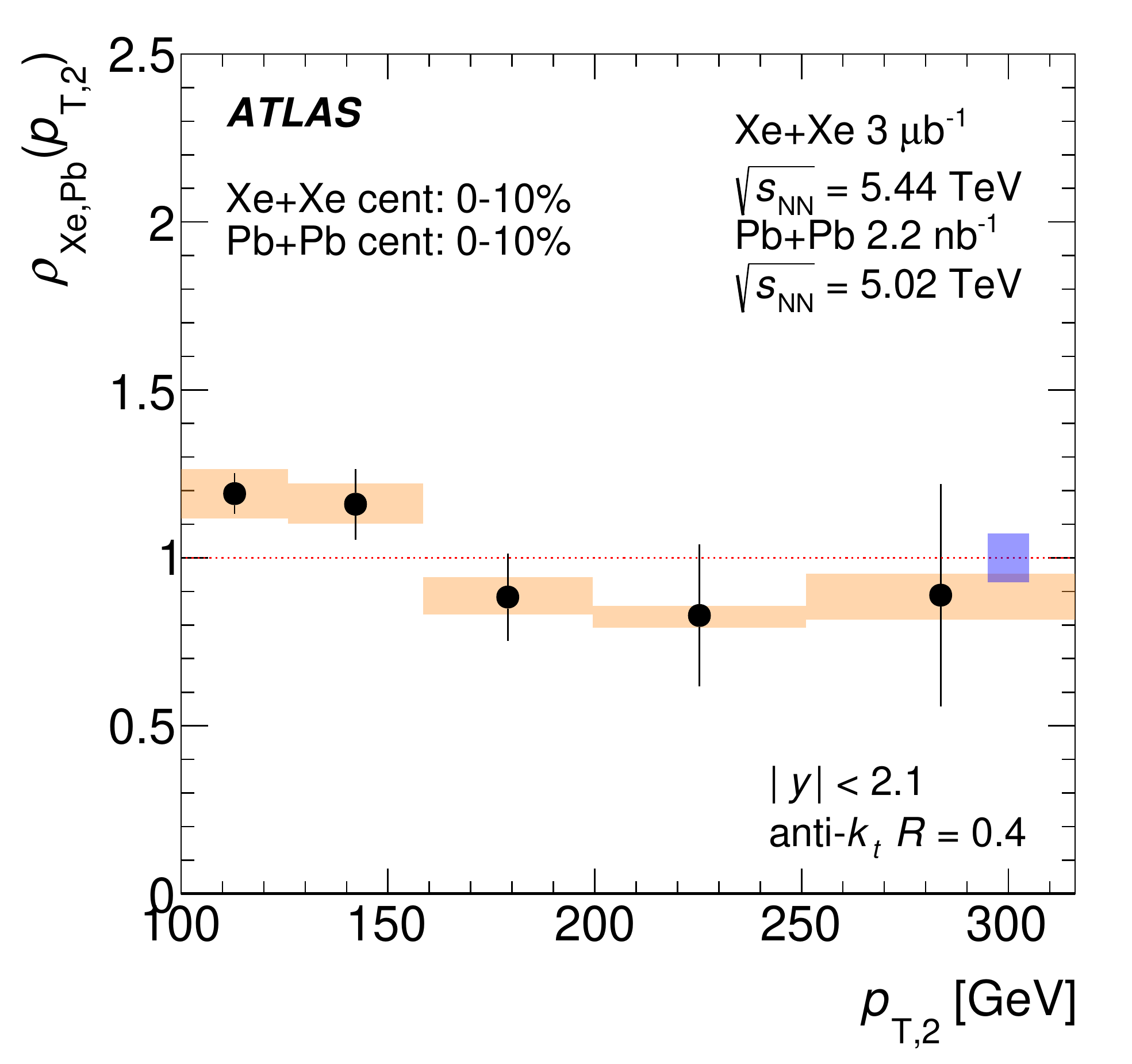}
\includegraphics[width=0.33\linewidth]{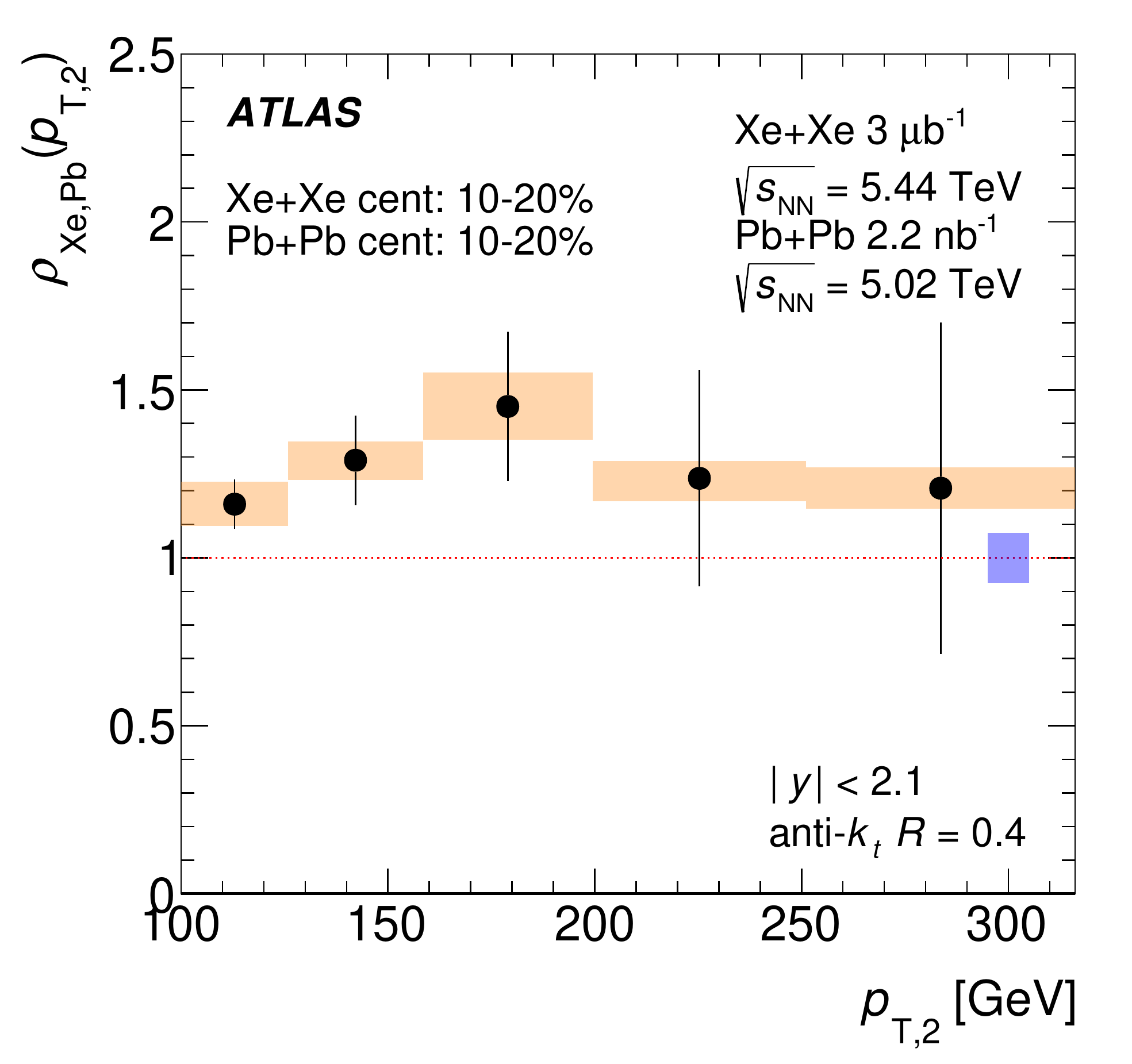}
\includegraphics[width=0.33\linewidth]{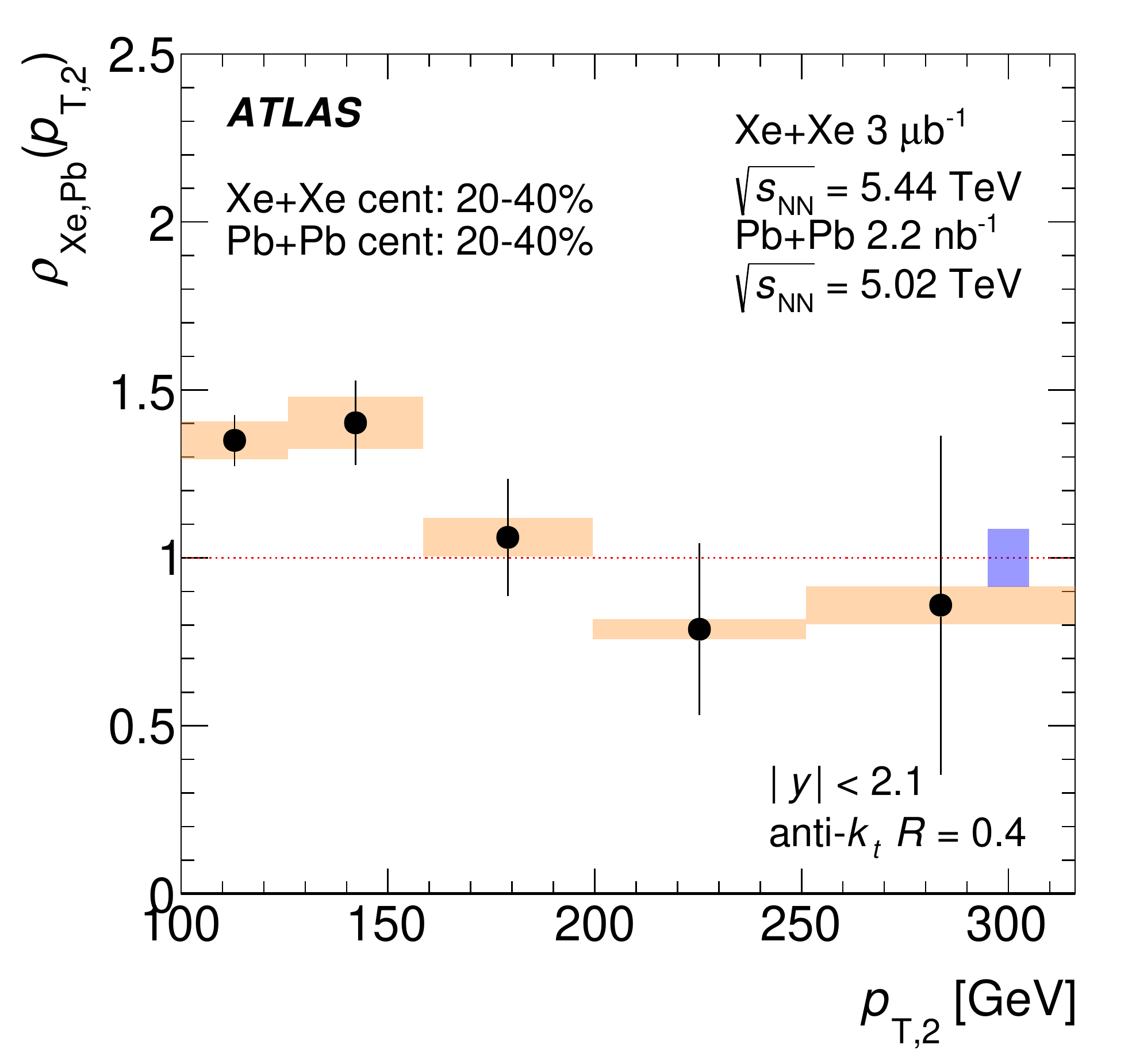}
}
\caption{The ratios of \xexe and \pbpb pair nuclear-modification factors, $\rhoab$, evaluated as a function of (upper plots) $\ptlead$ and (lower plots) $\ptsub$ in the same centrality intervals. Statistical and systematic uncertainties are represented by error bars and boxes, respectively. The box centered at unity represents the fractional systematic uncertainty on $\TAAavr$.}
\label{fig:cent:rab}
\end{figure}

\begin{figure}
\centering
\centerline{
\includegraphics[width=0.33\linewidth]{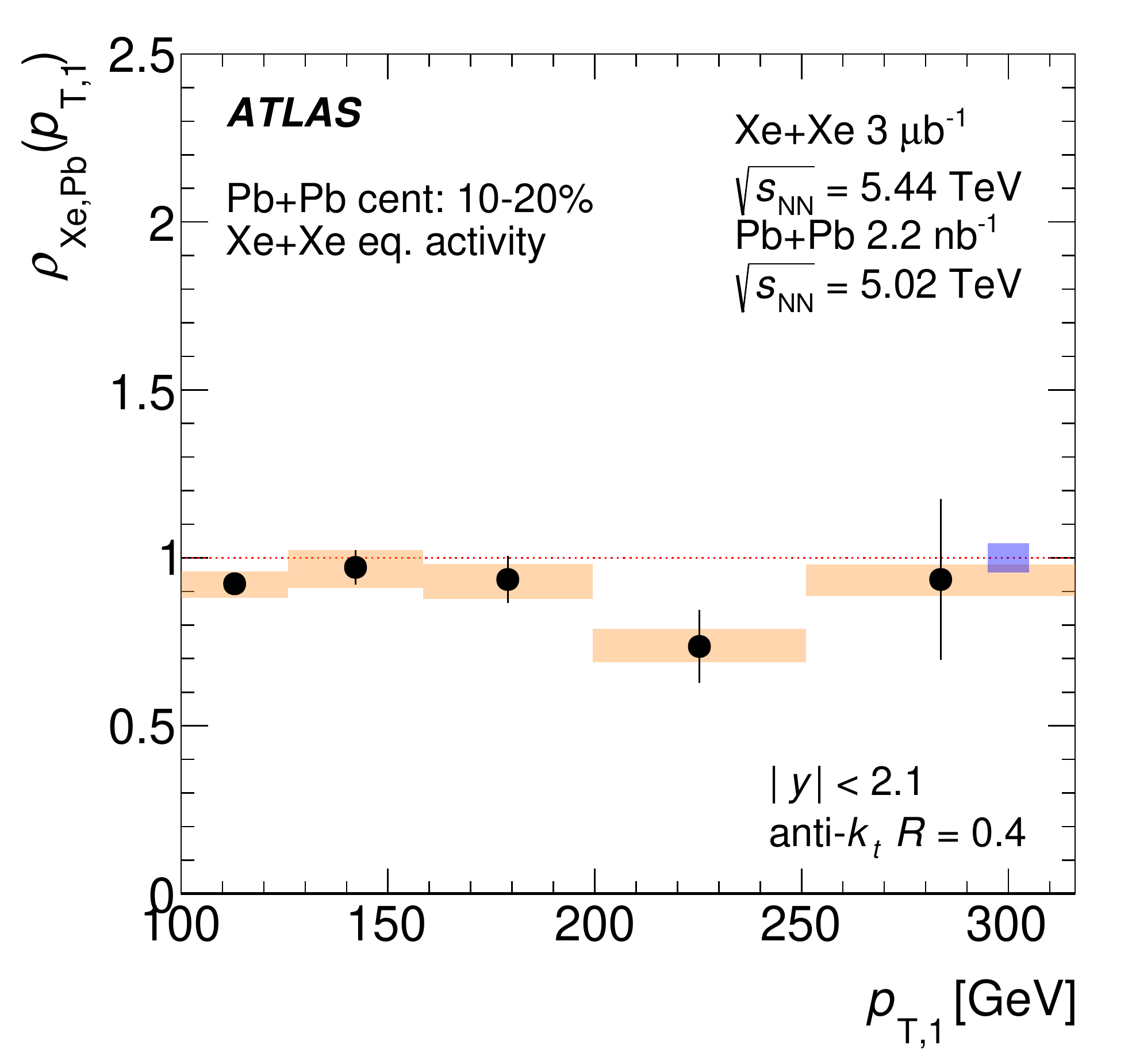}
\includegraphics[width=0.33\linewidth]{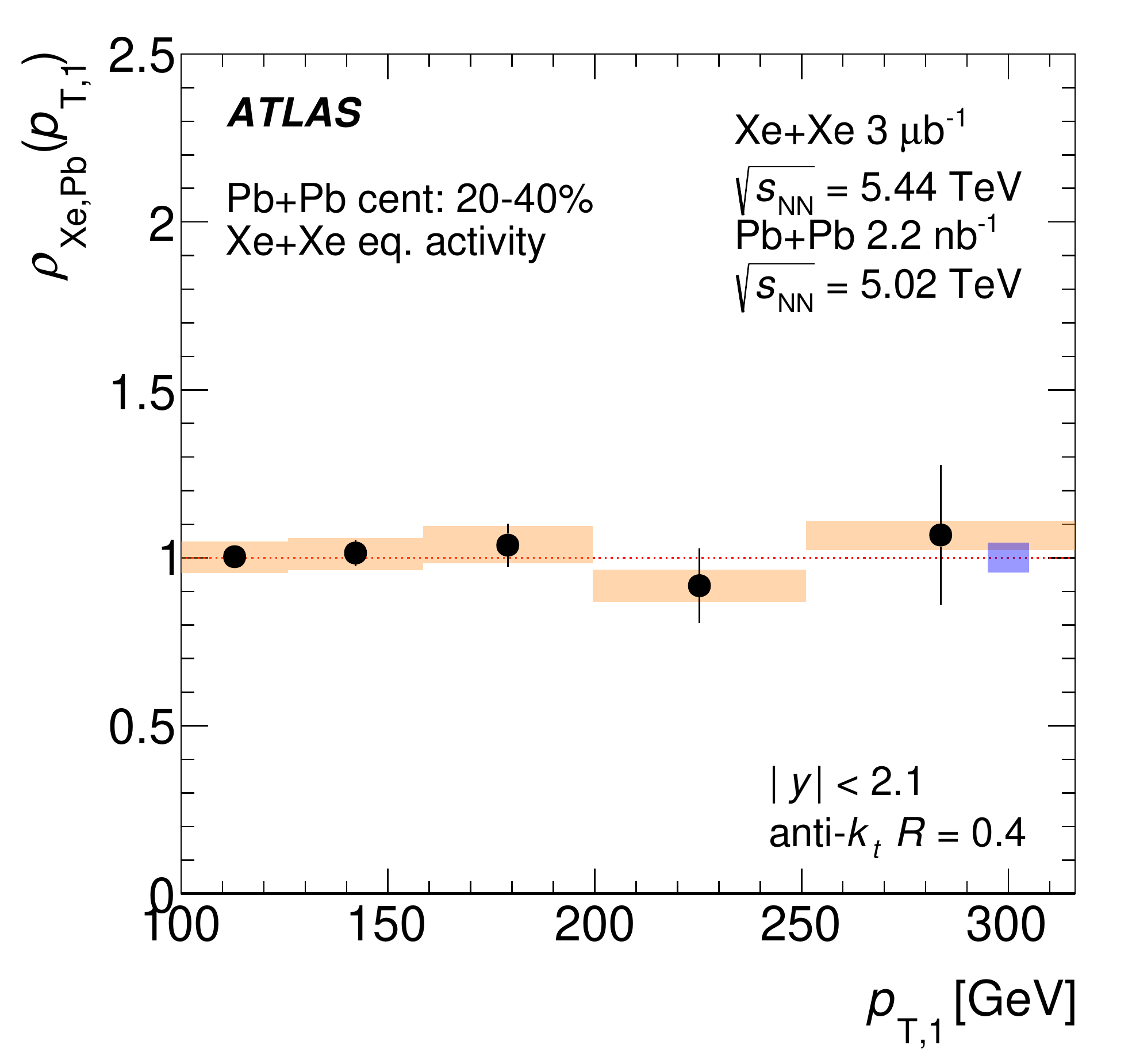}
\includegraphics[width=0.33\linewidth]{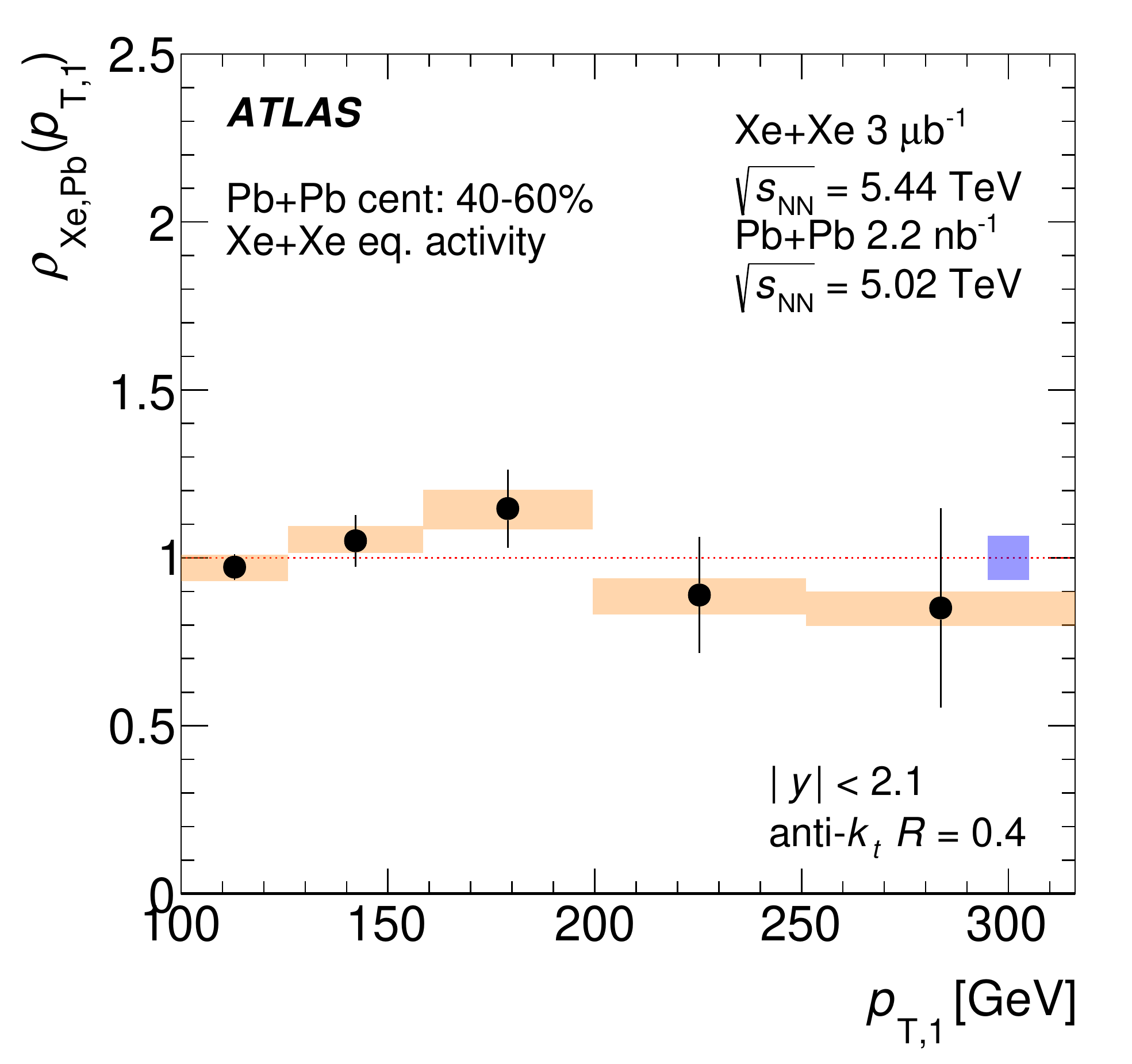}
}
\centerline{
\includegraphics[width=0.33\linewidth]{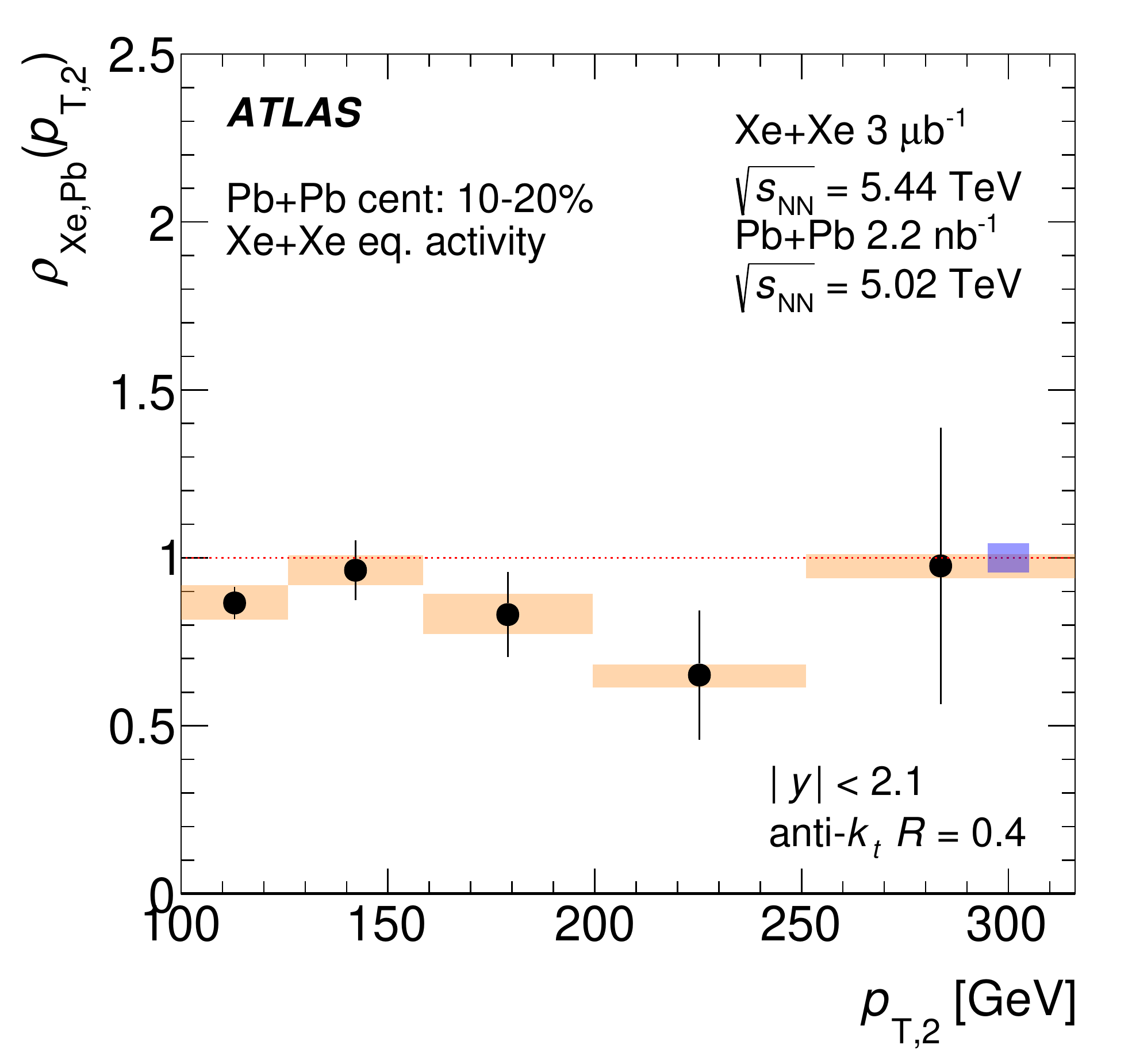}
\includegraphics[width=0.33\linewidth]{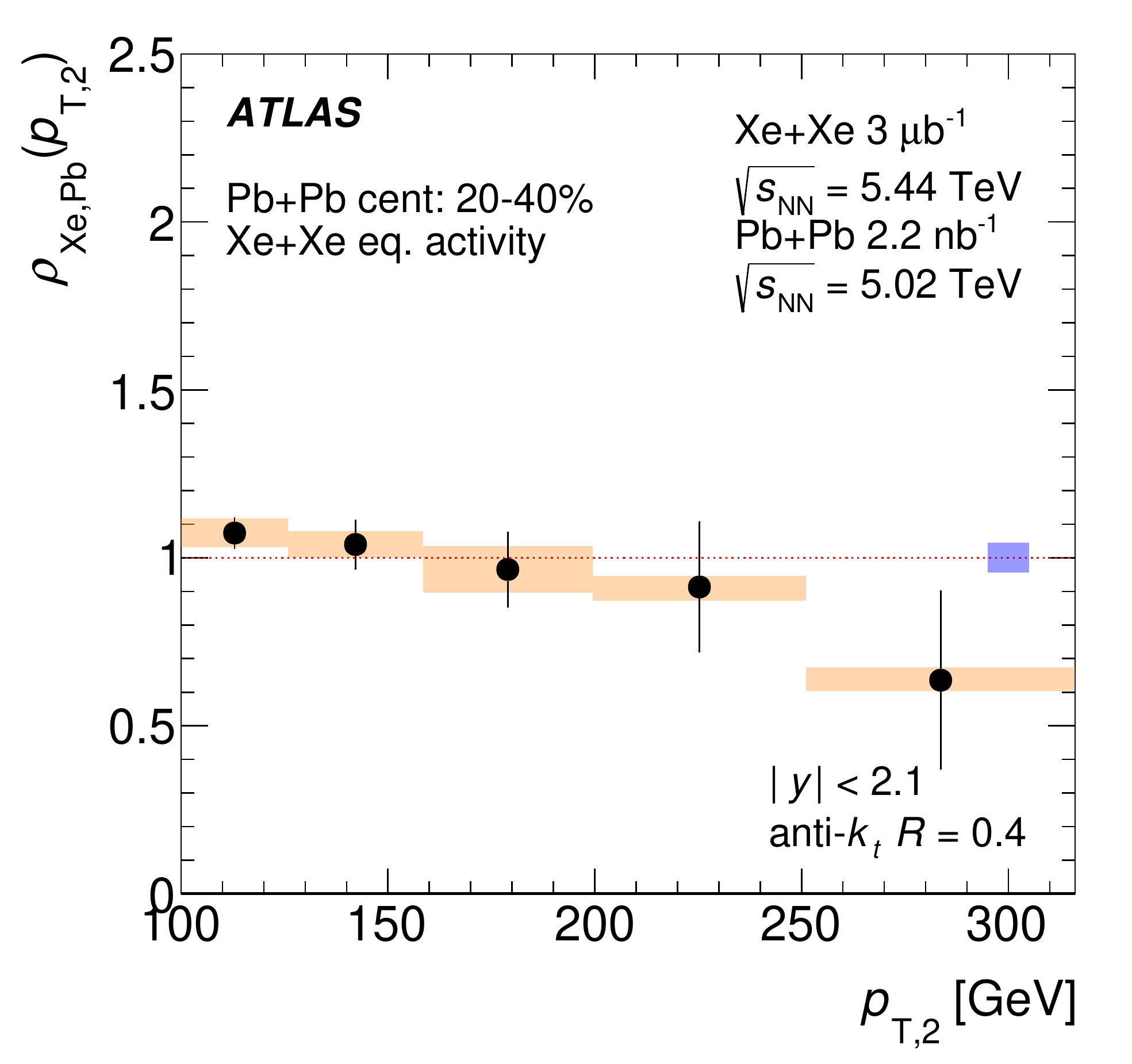}
\includegraphics[width=0.33\linewidth]{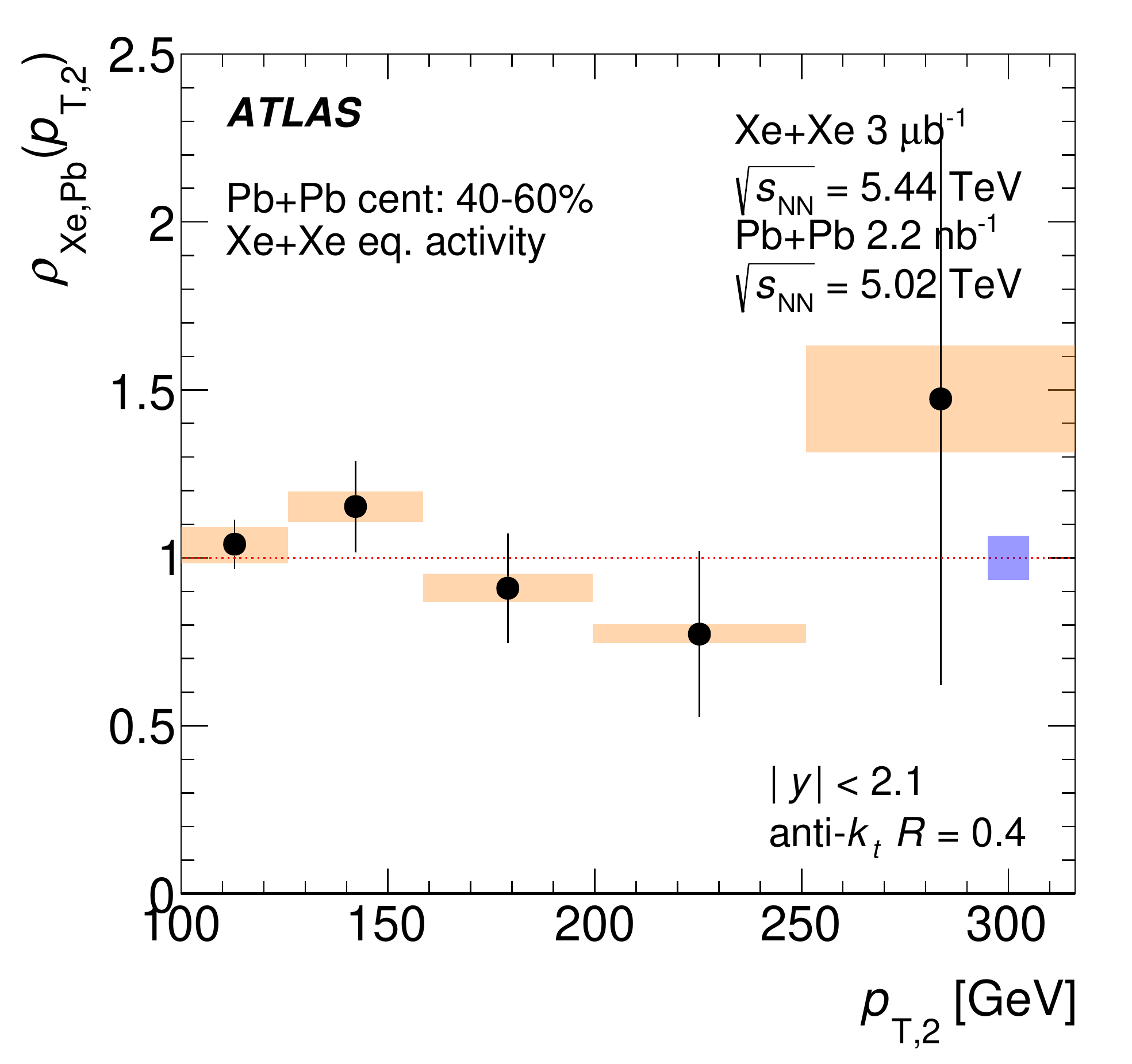}
}
\caption{The ratios of \xexe and \pbpb pair nuclear-modification factors, $\rhoab$, evaluated as a function of (upper plots) $\ptlead$ and (lower plots) $\ptsub$ in the same $\sumETFCal$ intervals (selecting equivalent event activity). Statistical and systematic uncertainties are represented by error bars and boxes, respectively. The box centered at unity represents the fractional systematic uncertainty on $\TAAavr$.}
\label{fig:fcal:rab}
\end{figure}

To characterize the differences between \xexe and \pbpb dijet suppression in a more quantitative way, the \xexe to \pbpb ratio of pair nuclear-modification factors, $\rhoab$, are evaluated as defined in Section~\ref{sec:analysis}. The $\rhoab(\ptlead)$ and $\rhoab(\ptsub)$ evaluated in the same \xexe and \pbpb centrality intervals are shown in Figure~\ref{fig:cent:rab}. The  $\rhoab$ values obtained are systematically larger than unity, typically by $10\%$ to $20\%$ depending on centrality. Figure~\ref{fig:fcal:rab} shows $\rhoab$ evaluated in the same event activity intervals.
In contrast to the centrality-based comparison, the $\rhoab$ values are consistent with unity within statistical and systematic uncertainties. This implies that the pair nuclear-modification factor in \xexe collisions at $\sqrtsnn = 5.44$~\TeV\ is consistent with the same quantity measured at $\sqrtsnn = 5.02$~\TeV\ in \pbpb collisions which suggests that the suppression of dijets does not differ in a significant way between \xexe and \pbpb collisions when measured in the same event activity intervals.
 
Despite consistency of \rhoab\ with the unity, we should still emphasize that any interpretation of the difference between the pair $\RAA$ evaluated as a function of $\ptlead$ and $\ptsub$ as the difference between the overall suppression of leading and subleading jets needs to take into account that the yields entering the pair $\RAA$ are conditional yields mutually dependent on kinematic selection criteria.
Consequently, any interpretation of $\rhoab(\ptlead), \rhoab(\ptsub)$ and $\raa^{\mathrm{pair}}(\ptlead), \raa^{\mathrm{pair}}(\ptsub)$ must be performed in the context of theoretical model predictions that directly follow the dijet definition and projection procedures used in this analysis.

\FloatBarrier
\section{Conclusions}
 
A measurement of per-pair normalized and absolutely normalized $\xj$ distributions of dijets in \xexe collisions at $\sqrtsnn~=~5.44$~\TeV\ using 3~$\mu$b$^{-1}$ of data collected with the ATLAS detector at the LHC is presented.
The per-pair normalized $\xj$ distributions indicate a higher relative rate of imbalanced dijets in central \xexe collisions compared with peripheral ones. The absolutely normalized $\xj$ distributions show that this feature arises predominantly from a depletion of the yields of more balanced dijets with $\xj$ values close to unity.
 
The results are compared with a measurement of dijets in \pbpb collisions at $\sqrtsnn=5.02$~\TeV\ by ATLAS\@.
The $\xj$ distributions are found to be consistent between \pbpb and \xexe collisions when compared in the same event activity intervals and after correcting the absolutely normalized distributions for the expected difference between the hard process cross-sections due to the different center-of-mass energies between \xexe and \pbpb collisions.
Furthermore, the differences between the dijet suppression in \xexe and \pbpb are quantified by the ratio of pair nuclear-modification factors, $\rhoab$, which are found to be consistent with unity when evaluated in the same event activity intervals.

These results should provide input for a better understanding of the role of path length, energy density, and fluctuations in the jet-energy loss in the QGP and add
a new input to the quantification of its system size dependence. In particular, $\rhoab$ consistent with unity can be used for predicting the magnitude of jet
quenching expected in future oxygen--oxygen collisions.
 
\section*{Acknowledgments}


We thank CERN for the very successful operation of the LHC, as well as the
support staff from our institutions without whom ATLAS could not be
operated efficiently.
 
We acknowledge the support of
ANPCyT, Argentina;
YerPhI, Armenia;
ARC, Australia;
BMWFW and FWF, Austria;
ANAS, Azerbaijan;
CNPq and FAPESP, Brazil;
NSERC, NRC and CFI, Canada;
CERN;
ANID, Chile;
CAS, MOST and NSFC, China;
Minciencias, Colombia;
MEYS CR, Czech Republic;
DNRF and DNSRC, Denmark;
IN2P3-CNRS and CEA-DRF/IRFU, France;
SRNSFG, Georgia;
BMBF, HGF and MPG, Germany;
GSRI, Greece;
RGC and Hong Kong SAR, China;
ISF and Benoziyo Center, Israel;
INFN, Italy;
MEXT and JSPS, Japan;
CNRST, Morocco;
NWO, Netherlands;
RCN, Norway;
MEiN, Poland;
FCT, Portugal;
MNE/IFA, Romania;
MESTD, Serbia;
MSSR, Slovakia;
ARRS and MIZ\v{S}, Slovenia;
DSI/NRF, South Africa;
MICINN, Spain;
SRC and Wallenberg Foundation, Sweden;
SERI, SNSF and Cantons of Bern and Geneva, Switzerland;
MOST, Taiwan;
TENMAK, T\"urkiye;
STFC, United Kingdom;
DOE and NSF, United States of America.
In addition, individual groups and members have received support from
BCKDF, CANARIE, Compute Canada and CRC, Canada;
PRIMUS 21/SCI/017 and UNCE SCI/013, Czech Republic;
COST, ERC, ERDF, Horizon 2020 and Marie Sk{\l}odowska-Curie Actions, European Union;
Investissements d'Avenir Labex, Investissements d'Avenir Idex and ANR, France;
DFG and AvH Foundation, Germany;
Herakleitos, Thales and Aristeia programmes co-financed by EU-ESF and the Greek NSRF, Greece;
BSF-NSF and MINERVA, Israel;
Norwegian Financial Mechanism 2014-2021, Norway;
NCN and NAWA, Poland;
La Caixa Banking Foundation, CERCA Programme Generalitat de Catalunya and PROMETEO and GenT Programmes Generalitat Valenciana, Spain;
G\"{o}ran Gustafssons Stiftelse, Sweden;
The Royal Society and Leverhulme Trust, United Kingdom.
 
The crucial computing support from all WLCG partners is acknowledged gratefully, in particular from CERN, the ATLAS Tier-1 facilities at TRIUMF (Canada), NDGF (Denmark, Norway, Sweden), CC-IN2P3 (France), KIT/GridKA (Germany), INFN-CNAF (Italy), NL-T1 (Netherlands), PIC (Spain), ASGC (Taiwan), RAL (UK) and BNL (USA), the Tier-2 facilities worldwide and large non-WLCG resource providers. Major contributors of computing resources are listed in Ref.~\cite{ATL-SOFT-PUB-2021-003}.


\clearpage
\printbibliography
 
\clearpage
\input{atlas_authlist}

\end{document}

%% file: atlas_authlist.tex
 
\begin{flushleft}
\hypersetup{urlcolor=black}
{\Large The ATLAS Collaboration}

\bigskip

\AtlasOrcid[0000-0002-6665-4934]{G.~Aad}$^\textrm{\scriptsize 102}$,
\AtlasOrcid[0000-0002-5888-2734]{B.~Abbott}$^\textrm{\scriptsize 120}$,
\AtlasOrcid[0000-0002-1002-1652]{K.~Abeling}$^\textrm{\scriptsize 55}$,
\AtlasOrcid[0000-0002-8496-9294]{S.H.~Abidi}$^\textrm{\scriptsize 29}$,
\AtlasOrcid[0000-0002-9987-2292]{A.~Aboulhorma}$^\textrm{\scriptsize 35e}$,
\AtlasOrcid[0000-0001-5329-6640]{H.~Abramowicz}$^\textrm{\scriptsize 151}$,
\AtlasOrcid[0000-0002-1599-2896]{H.~Abreu}$^\textrm{\scriptsize 150}$,
\AtlasOrcid[0000-0003-0403-3697]{Y.~Abulaiti}$^\textrm{\scriptsize 117}$,
\AtlasOrcid[0000-0003-0762-7204]{A.C.~Abusleme~Hoffman}$^\textrm{\scriptsize 137a}$,
\AtlasOrcid[0000-0002-8588-9157]{B.S.~Acharya}$^\textrm{\scriptsize 69a,69b,p}$
% ...
\bigskip
\\

% ...
$^{29}$Physics Department, Brookhaven National Laboratory, Upton NY; United States of America.\\
% ...
$^{35}$$^{(a)}$Facult\'e des Sciences Ain Chock, R\'eseau Universitaire de Physique des Hautes Energies - Universit\'e Hassan II, Casablanca;$^{(b)}$Facult\'{e} des Sciences, Universit\'{e} Ibn-Tofail, K\'{e}nitra;$^{(c)}$Facult\'e des Sciences Semlalia, Universit\'e Cadi Ayyad, LPHEA-Marrakech;$^{(d)}$LPMR, Facult\'e des Sciences, Universit\'e Mohamed Premier, Oujda;$^{(e)}$Facult\'e des sciences, Universit\'e Mohammed V, Rabat;$^{(f)}$Institute of Applied Physics, Mohammed VI Polytechnic University, Ben Guerir; Morocco.\\
% ...
$^{55}$II. Physikalisches Institut, Georg-August-Universit\"{a}t G\"ottingen, G\"ottingen; Germany.\\
% ...
$^{69}$$^{(a)}$INFN Gruppo Collegato di Udine, Sezione di Trieste, Udine;$^{(b)}$ICTP, Trieste;$^{(c)}$Dipartimento Politecnico di Ingegneria e Architettura, Universit\`a di Udine, Udine; Italy.\\
% ...
$^{102}$CPPM, Aix-Marseille Universit\'e, CNRS/IN2P3, Marseille; France.\\
% ...
$^{117}$Department of Physics, New York University, New York NY; United States of America.\\
% ...
$^{120}$Homer L. Dodge Department of Physics and Astronomy, University of Oklahoma, Norman OK; United States of America.\\
% ...
$^{137}$$^{(a)}$Departamento de F\'isica, Pontificia Universidad Cat\'olica de Chile, Santiago;$^{(b)}$Millennium Institute for Subatomic physics at high energy frontier (SAPHIR), Santiago;$^{(c)}$Instituto de Investigaci\'on Multidisciplinario en Ciencia y Tecnolog\'ia, y Departamento de F\'isica, Universidad de La Serena;$^{(d)}$Universidad Andres Bello, Department of Physics, Santiago;$^{(e)}$Instituto de Alta Investigaci\'on, Universidad de Tarapac\'a, Arica;$^{(f)}$Departamento de F\'isica, Universidad T\'ecnica Federico Santa Mar\'ia, Valpara\'iso; Chile.\\
% ...
$^{150}$Department of Physics, Technion, Israel Institute of Technology, Haifa; Israel.\\
$^{151}$Raymond and Beverly Sackler School of Physics and Astronomy, Tel Aviv University, Tel Aviv; Israel.\\
% ...
$^{a}$ Also Affiliated with an institute covered by a cooperation agreement with CERN.\\
$^{b}$ Also at An-Najah National University, Nablus; Palestine.\\
$^{c}$ Also at Borough of Manhattan Community College, City University of New York, New York NY; United States of America.\\
$^{d}$ Also at Bruno Kessler Foundation, Trento; Italy.\\
$^{e}$ Also at Center for High Energy Physics, Peking University; China.\\
$^{f}$ Also at Center for Interdisciplinary Research and Innovation (CIRI-AUTH), Thessaloniki ; Greece.\\
% ...
$^{p}$ Also at Department of Physics, King's College London, London; United Kingdom.\\
% ...

\end{flushleft}
